\documentclass[11pt,a4paper]{article}
\pdfoutput=1
\usepackage{jcappub}

\usepackage{epsfig,psfrag,graphics,verbatim}
\usepackage{graphicx}

\usepackage{dcolumn}
\usepackage{bm}
\usepackage{epsfig}
\usepackage{amsmath,amssymb}
\usepackage{psfig}
\usepackage{ulem}
\usepackage{natbib}

\def \apj  {ApJ}

\def \prd {Phy. Rev. D}
\def \mnras {MNRAS}
\def \chisq  {\ifmmode  \chi^2   \else  $\chi^2$  \fi}  
\def \spose#1{\hbox  to 0pt{#1\hss}}  
\def \lta{\mathrel{\spose{\lower 3pt\hbox{$\sim$}}\raise  2.0pt\hbox{$<$}}}
\def \gta{\mathrel{\spose{\lower  3pt\hbox{$\sim$}}\raise 2.0pt\hbox{$>$}}}

\title{Hints on the Nature of Dark Matter from the Properties of Milky Way Satellites}

\author[a]{Donnino Anderhalden}
\author[b]{Aurel Schneider}
\author[c]{Andrea V. Macci\`o}
\author[a]{\\Juerg Diemand}
\author[d]{Gianfranco Bertone}

\affiliation[a]{Institute for Theoretical Physics, University of Z\"urich, Winterthurerstrasse 190, 8057 Z\"urich, Switzerland}
\affiliation[b]{Department of Physics \& Astronomy, University of Sussex, Brighton, BN1 9QH, United Kingdom}
\affiliation[c]{Max-Planck-Institute for Astronomy, K\"onigstuhl 17, 69117 Heidelberg, Germany}
\affiliation[d]{GRAPPA Institute, University of Amsterdam, Science Park 904, 1090 GL Amsterdam, Netherlands}
\emailAdd{donninoa@physik.uzh.ch}
\emailAdd{aurel.schneider@sussex.ac.uk}
\emailAdd{maccio@mpia.de}
\emailAdd{diemand@physik.uzh.ch}
\emailAdd{gf.bertone@gmail.com}

\abstract{The nature of dark matter is still unknown and one of the most fundamental scientific mysteries. Although successfully describing large scales, the standard cold dark matter model (CDM) exhibits possible shortcomings on galactic and sub-galactic scales. It is exactly at these highly non-linear scales where strong astrophysical constraints can be set on the nature of the dark matter particle. While observations of the Lyman-$\alpha$ forest probe the matter power spectrum in the mildly non-linear regime, satellite galaxies of the Milky Way provide an excellent laboratory as a test of the underlying cosmology on much smaller scales. Here we present results from a set of high resolution simulations of a Milky Way sized dark matter halo in eight distinct cosmologies: CDM, warm dark matter (WDM) with a particle mass of 2 keV and six different cold plus warm dark matter (C+WDM) models, varying the fraction, $f_{\rm wdm}$, and the mass, $m_{\rm wdm}$, of the warm component. We used three different observational tests based on Milky Way satellite observations: the total satellite abundance, their radial distribution and their mass profile. We show that the requirement of simultaneously satisfying all three constraints sets very strong limits on the nature of dark matter. This shows the power of a multi-dimensional small scale approach in ruling out models which would be still allowed by large scale observations.}


\begin{document}
\maketitle

\section{Introduction}
\label{introduction}
On large scales cold dark matter (CDM) predictions agree very well with galaxy clustering surveys such as the Sloan Digital Sky Survey (SDSS) \cite[][]{SDSScollaboration} and 6dF\footnote{The SDSS and 6dFGS databases can be found on {\it http://www-wfau.roe.ac.uk/6dFGS/} and {\it http://www.sdss.org/} respectively.} \cite[][]{6dFcollaboration2009}. On galactic and sub-galactic scales some tension between observations and numerical CDM simulations appears and hence there are a some small scale challenges for the CDM paradigm. In the following we want to emphasize three issues discussed in the literature.
The first one is the so called core/cusp problem: the estimated mass profiles of dark matter haloes inferred from observations are mostly consistent with constant density cores \cite[e.g.][]{Swatersetal2003,Gentileetal2009,Strigarietal2010,Walker2011, Agnello2012}, whereas simulations of CDM haloes tend to produce cuspy inner density profiles \cite[e.g.][]{Moore94,Diemandetal2005,maccio2012}. The second concern is the overabundance of small substructure in Milky Way sized dark matter halos when compared with observed ultra-faint satellite galaxies \cite[][]{Klypinetal1999,Mooreetal1999c,Madauetal2008,Munozetal2009,Maccioetal2010,Rashkovetal2012}. A third issue is that several of the most massive CDM subhalos seem to be too concentrated in order to host the brightest observed satellite galaxies. The $\Lambda$CDM model predicts the existence of large, dark galactic subhalos, which seem to be "too big to fail" to form a dwarf galaxy \cite[e.g.][]{Boylan-Kolchinetal2011,Lovelletal2012, Rashkovetal2012, Tollerudetal2012}.

Even though several baryonic solutions have been proposed to solve at least some of the inconsistencies mentioned above \cite[e.g.][]{WadepuhlSpringel,Parryetal2012,Governatoetal2012}, it is worth exploring alternative, non-standard scenarios.
If the dark matter particle is much lighter (with a mass in the keV region) than its CDM counterpart, it decouples from the hot plasma in the early Universe when it is still relativistic. Hence, dark matter particles are able to free-stream until relatively late times, washing out small density perturbations and suppress the formation of structure below a characteristic scale (which is denoted as the free streaming scale, $\lambda_{\rm fs}$). Such dark matter models are named Warm Dark Matter (WDM) \cite[e.g.][and references therein]{Bodeetal2001,Schneideretal2012}. Typical candidates are gravitino \cite[][]{Ellisetal1984, Moroietal1993,Gorbunovetal2008} and sterile neutrino like particles \cite[e.g.][]{DodelsonWidrow1994,ShaposhnikovTkachev2006,Shaposhnikov2007}.

Another viable solution is to assume that the total dark matter density is made of a mixture of cold plus warm dark matter (C+WDM) \cite[][]{boyarsky2009,Boyarskyetal2009a,Boyarskyetal2009b,MaccioMixed}. In addition to the mass of the warm particle, these class of models are also characterised by the fraction of warm to cold dark matter, i.e.
\begin{equation}\label{wdmfraction}
f_{\rm W} = \frac{\Omega_{\rm WDM}}{\Omega_{\rm WDM} + \Omega_{\rm CDM}} = \frac{\Omega_{\rm WDM}}{\Omega_{\rm DM}}.
\end{equation}
Although afflicted with large uncertainties, currently, the best way to distinguish between CDM and non-standard models is to analyse data from the Lyman-$\alpha$ forest \cite[for a review see][]{Meiksin2007}. As we will show in this work, a dark matter model, even when not ruled out by Ly-$\alpha$ physics, still has to pass a series of test on much smaller scales: in order to be hosts of the brightest Milky Way satellites, subhalos in a galaxy-sized dark matter halo should be in line with observations in terms of their abundance, radial distribution and mass profiles. 

This paper is organised as follows. Section \ref{simulations} provides a brief summary of the techniques we used for computing the transfer function for mixed dark matter models as well as an overview of our numerical simulations (see also \cite[][]{Anderhaldenetal2012} for more details). Section \ref{results} shows the results of our subhalo analysis: subhalo abundance, radial distribution and mass profiles are compared with observed satellite galaxies. A discussion and conclusion is  presented in section \ref{discussion}.

\section{Cosmological Models and N-body Simulations}\label{simulations}
In this section we briefly introduce the cosmological models used in this work, as well as the simulation details. Besides the conventional cold dark matter paradigm, we run a warm dark matter cosmology with a particle mass of 2 keV as well as six different cold plus warm dark matter (C+WDM) cosmologies. It should be emphasized that all non-standard models except two (f20 - 0.1keV \& f50 - 0.3keV) lie within the two sigma range of validity for Ly-$\alpha$ forest constraints \cite[see e.g.][for recent constraints]{Boyarskyetal2009a}.

A generic C+WDM model is described by the mass of the warm particle, $m_{\rm WDM}$, and the fraction of warm to cold dark matter, $f_{\rm W}$ (see eq. \eqref{wdmfraction}). We have chosen a set of six mixed dark matter models by scanning through the $f_{\rm W}$ - $m_{\rm WDM}$ parameter space. In order to isolate effects coming from either the fraction or the mass, in some of the models only one of the parameters is changed. Unlike the numerical C+WDM realisations in \cite{Boyarskyetal2009a,MaccioMixed}, where every particle in the simulation represents the mixture of both species, our two particle species approach allows us to distinguish the cold and warm components (similar as for example in SPH simulations containing baryons). The initial conditions are set up using two cosmological boxes, one containing only cold particles, the other only warm particles, and merging them with an offset of half a grid cell in each dimension.
\begin{figure}
\begin{center}
\includegraphics[scale=0.37]{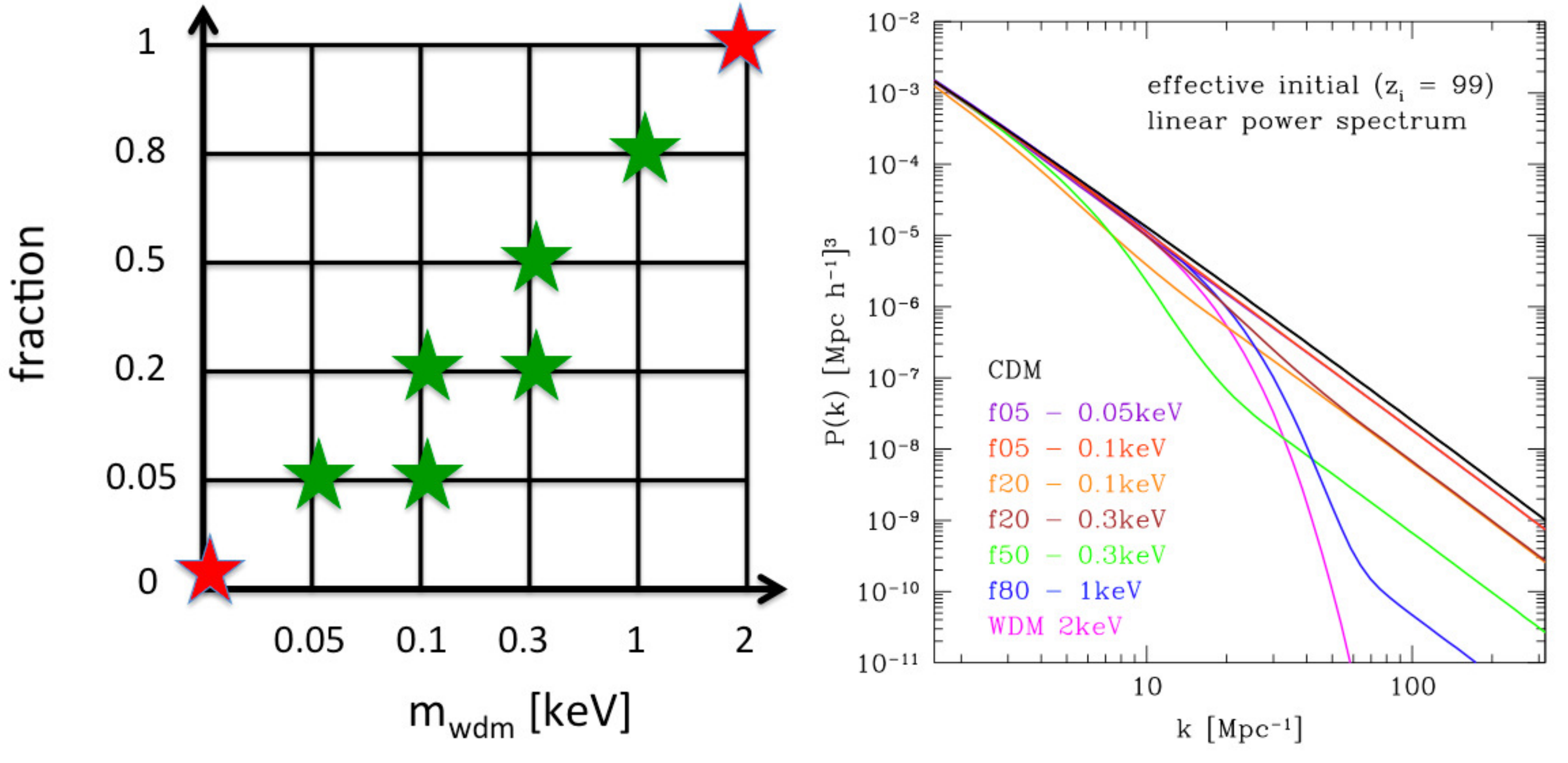}
\caption{Left panel: set of N-body simulations in the $f_{\rm W}$ - $m_{\rm WDM}$ parameter space. Green stars denote mixed dark matter simulations whereas red stars are pure cold ($f_{\rm W} = 0$) and pure warm ($f_{\rm W} = 1$) dark matter simulations respectively. Right panel: matter power spectrum of all simulated models with increasing $f_{\rm W}$ from top to bottom (at large $k$).}
\label{gridpower}
\end{center}
\end{figure}
A detailed description of the numerical methods is provided in section 2 of \cite{Anderhaldenetal2012}, hence in the following we just briefly sketch the simulation procedure and the implementation of the corresponding transfer functions.
\begin{table}
\caption{Details of the simulations. $R_{200}$, $N_{200}$ and $M_{200}$ are measured with respect to 200 times the mean matter density.}
\label{simulationoverview}
\begin{center}
\begin{tabular}{lcccccc}
\hline
\hline 
Label & m$_{\text{WDM}}$ & $\bar{f}_{\rm W}$ & $\sigma_{\text{th}}^{z=99}$ & $R_{200}$ & $N_{200}$ & $M_{200}$ \\ 
& [keV] & [\%] & [km s$^{-1}$] & [kpc] & [$\times 10^6$] & [$10^{12}$ M$_{\odot}$] \\
\hline\hline
CDM & - & 0 & 0 & 367 & 9.88 & 1.36\\
\hline
f05 - 0.05keV & 0.05 & 5 & 82.1 & 362 & 8.98 & 1.24\\
f05 - 0.1keV & 0.1 & 5 & 32.6 & 368 & 9.46 & 1.31\\
f20 - 0.1keV & 0.1 & 20 & 51.7 & 365 & 8.49 & 1.17\\ 
f20 - 0.3keV & 0.3 & 20 & 11.9 & 368 & 9.20 & 1.27 \\ 
f50 - 0.3keV & 0.3 & 50 & 16.2 & 367 & 9.01 & 1.24\\
f80 - 1keV & 1.0 & 80 & 3.8 & 368 & 8.69 & 1.20\\
\hline
WDM & 2.0 & 100 & 1.6 & 367 & 9.04 & 1.25 \\
\hline\hline
\end{tabular}
\end{center}
\end{table}
At first, we run large scale simulation boxes with length of 40 Mpc with a total particle number of $2\times 256^3$. The transfer functions of the non-CDM models, representing small scale suppression, can be written as
\begin{equation}
T^2(k) = \frac{P_{\alpha}(k)}{P_{\rm CDM}(k)},\qquad \alpha = \{\text{WDM, C+WDM}\}.
\end{equation}
For the pure WDM run, we used the fitting formula suggested by \cite{Vieletal2005},
\begin{equation}
T_{\text{WDM}}(k)^2 = \Big(1 + \big(\alpha k \big)^{2\nu} \Big)^{-10/\nu},
\end{equation}
where $\nu=1.12$ and $\alpha$ is a function of the warm dark matter particle mass. In the mixed dark matter case, we made the simplistic assumption that the universe is filled with a dissipationless, coupled cold plus warm dark matter fluid and numerically solved the basic hydrodynamical equations of Newtonian physics. Linear theory then allows us to connect density contrasts between two epochs (i.e. matter-radiation equality and the starting redshift of the simulations) via a mixed dark matter transfer function $T(k)$ \cite[see][for details]{Anderhaldenetal2012}. The right panel in figure \ref{gridpower} shows the linear power spectrum at the starting redshift of our simulations $z_{\rm IC}=99$. Unlike the pure WDM case, where the linear power rapidly drops to zero, the linear power spectrum in mixed dark matter models stabilizes due to the presence of a cold component and approaches a constant plateau with a characteristic height only dependent on the fraction $f_{\rm W}$. All simulations have been performed with the parallel treecode {\sc pkdgrav}, written by Joachim Stadel and Thomas Quinn \cite{Stadel2001}. Initial conditions are generated with a parallel version of the GRAFIC package \cite{Bertschinger2001} and are based on the cosmological parameters taken from WMAP7 \cite{Komatsuetal2011}: $\sigma_8=0.8$, $h=0.7$, $\Omega_{\rm dm}=0.227$, $\Omega_{\rm b}=0.046$, $\Omega_{\Lambda}=0.727$ and $n_{\rm s}=0.961$. For the sake of completeness, figure \ref{gridpower} shows the resulting grid of high resolution N-body simulations.

In a second step we selected an isolated and Milky Way sized ($\sim 10^{12} M_{\odot}$) halo at redshift $z=0$ and re-simulated a Lagrangian region of $3 \times R_{200}$ of the original object at $8^3$ times better mass resolution. This leads to a particle mass of $m_p=1.38\times 10^5 M_{\odot}$ and a gravitational softening of 355 pc in the refined region.

\begin{figure}
\begin{center}
\includegraphics[scale=0.1338]{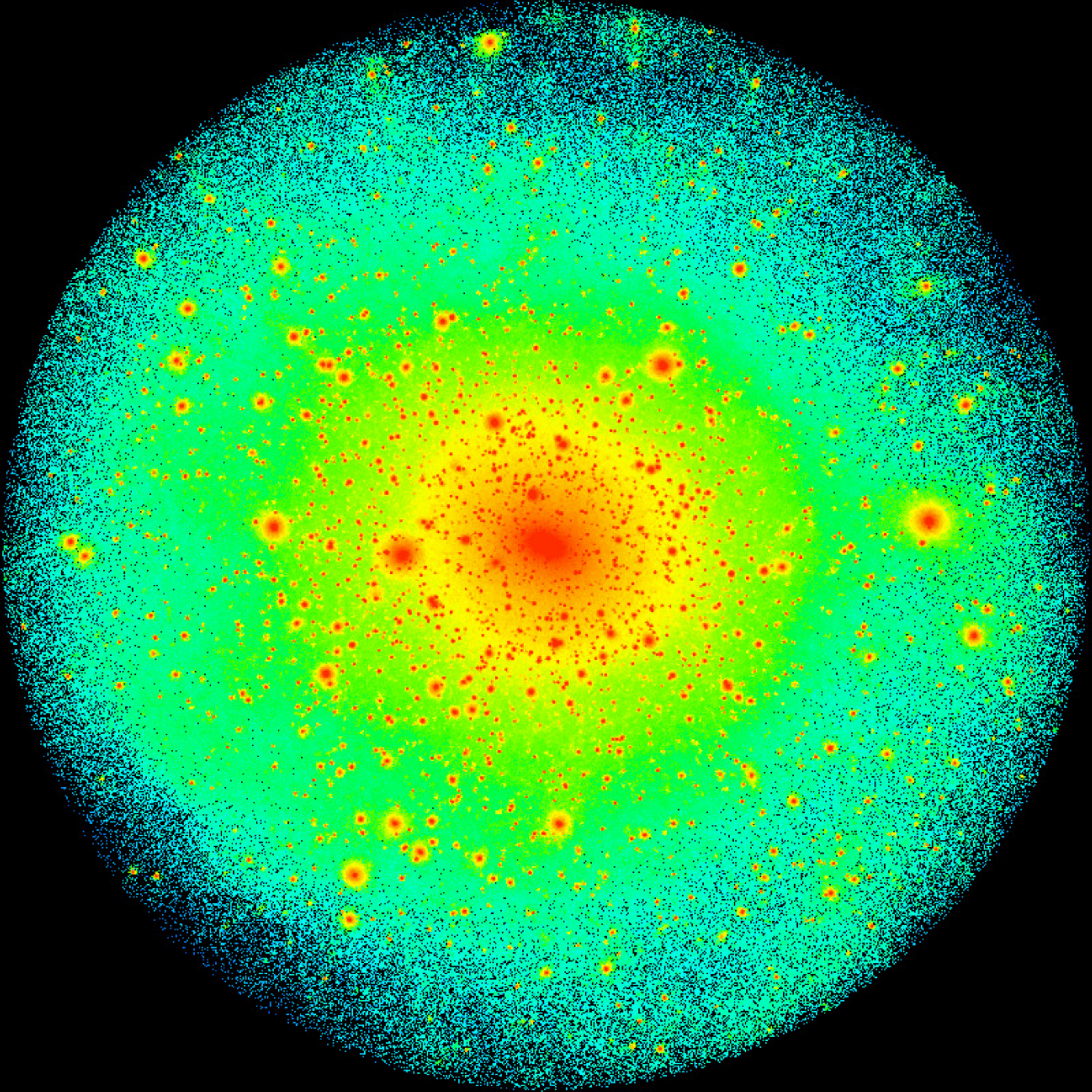}
\includegraphics[scale=0.1338]{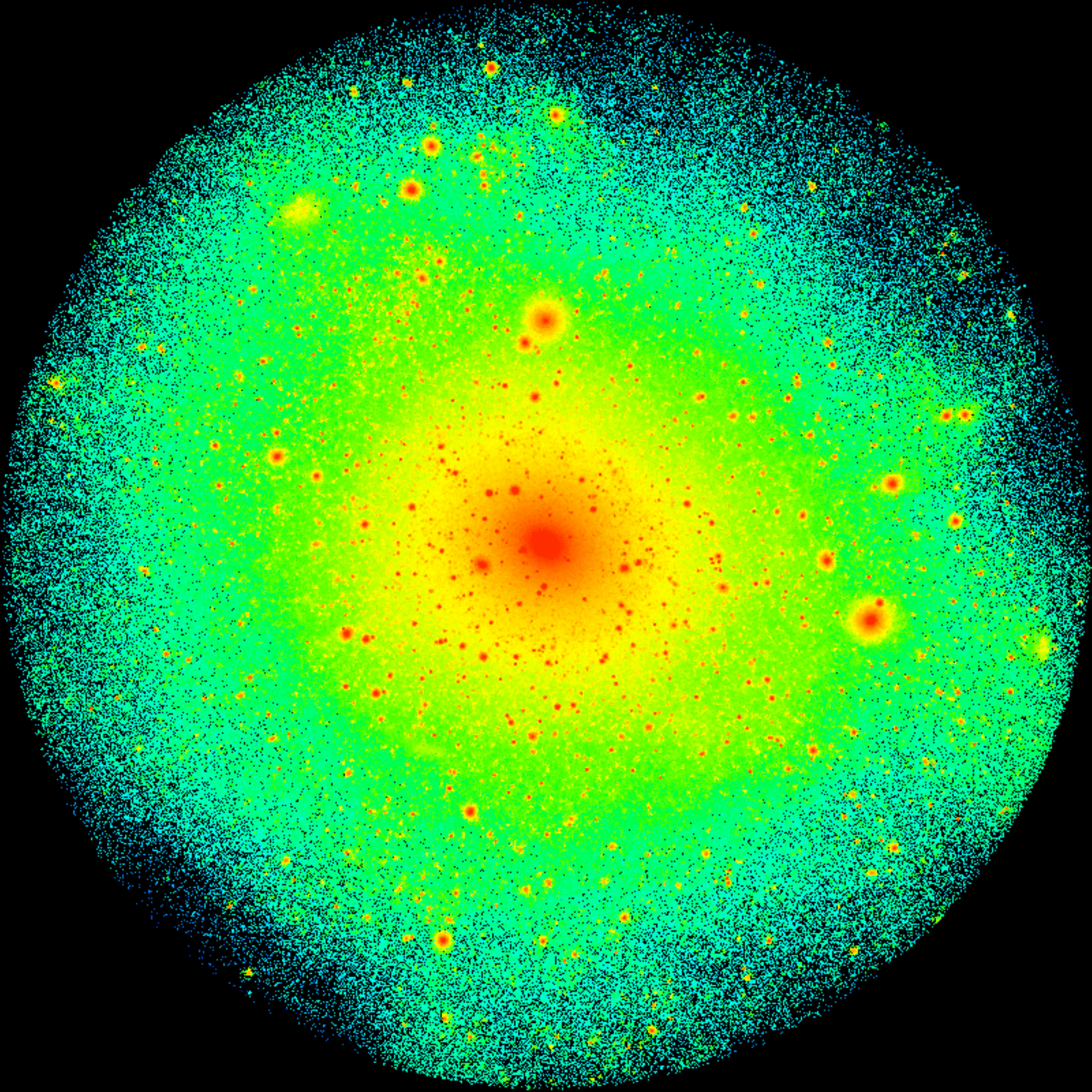}
\includegraphics[scale=0.1338]{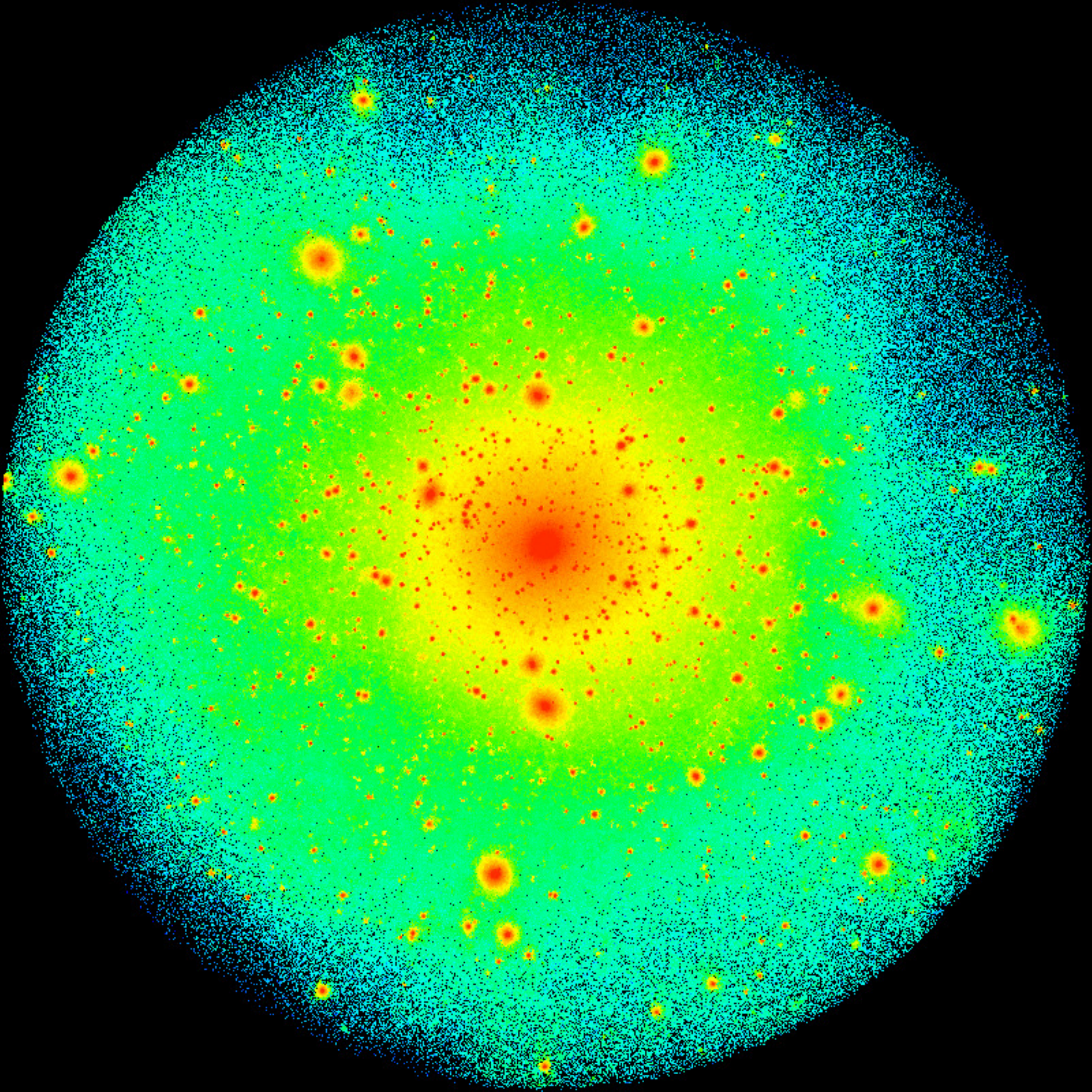}
\includegraphics[scale=0.1338]{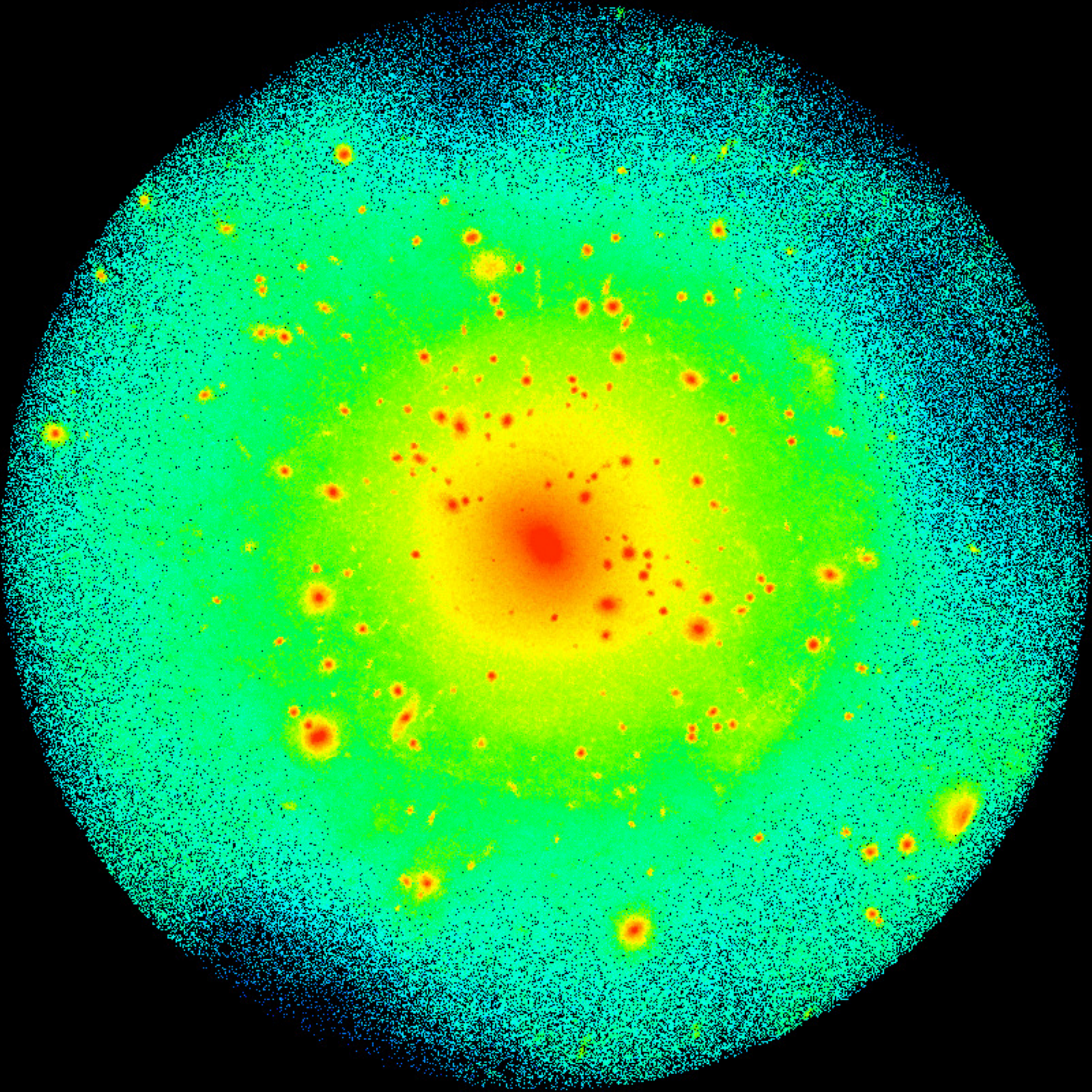}
\newline
\includegraphics[scale=0.1338]{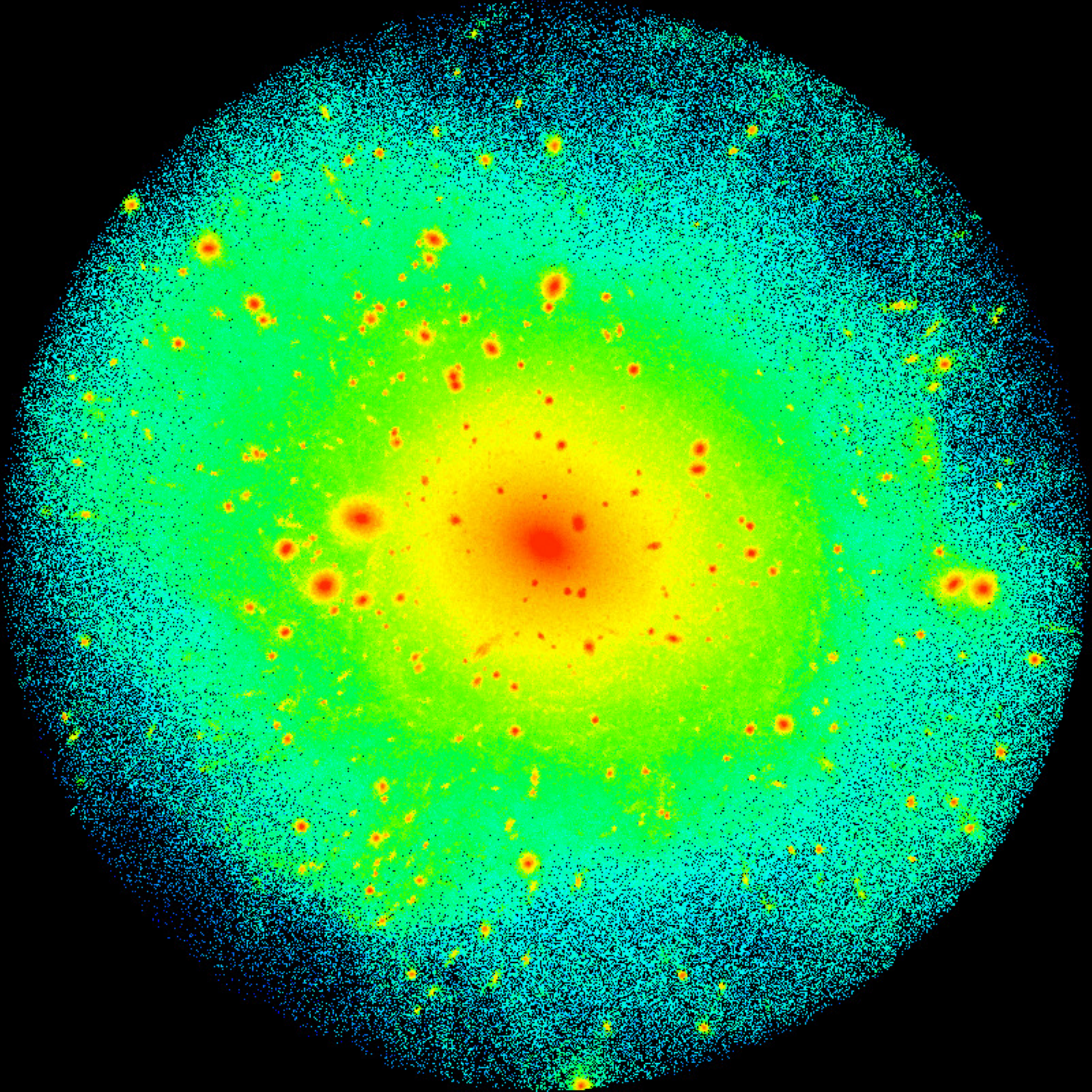}
\includegraphics[scale=0.1338]{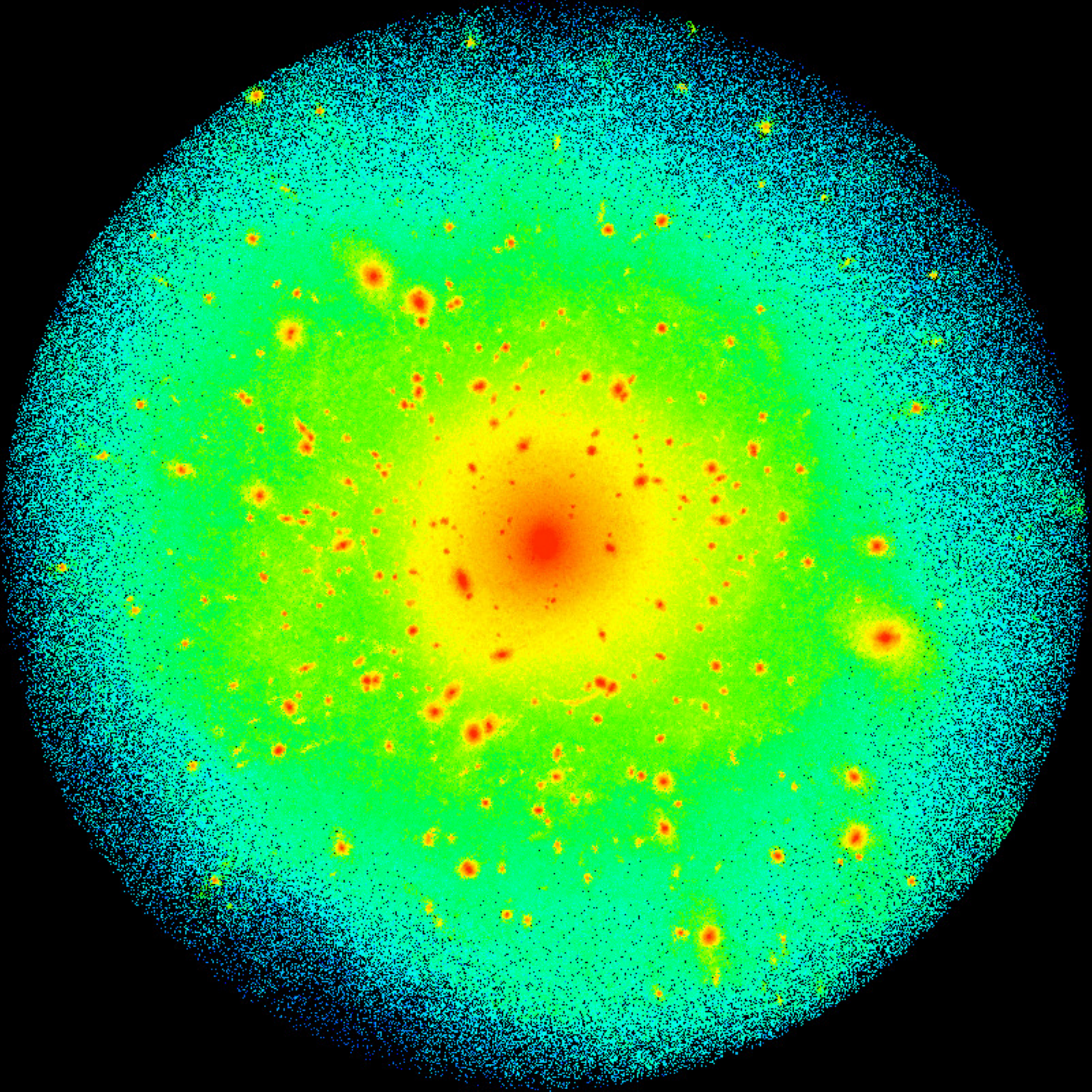}
\includegraphics[scale=0.1338]{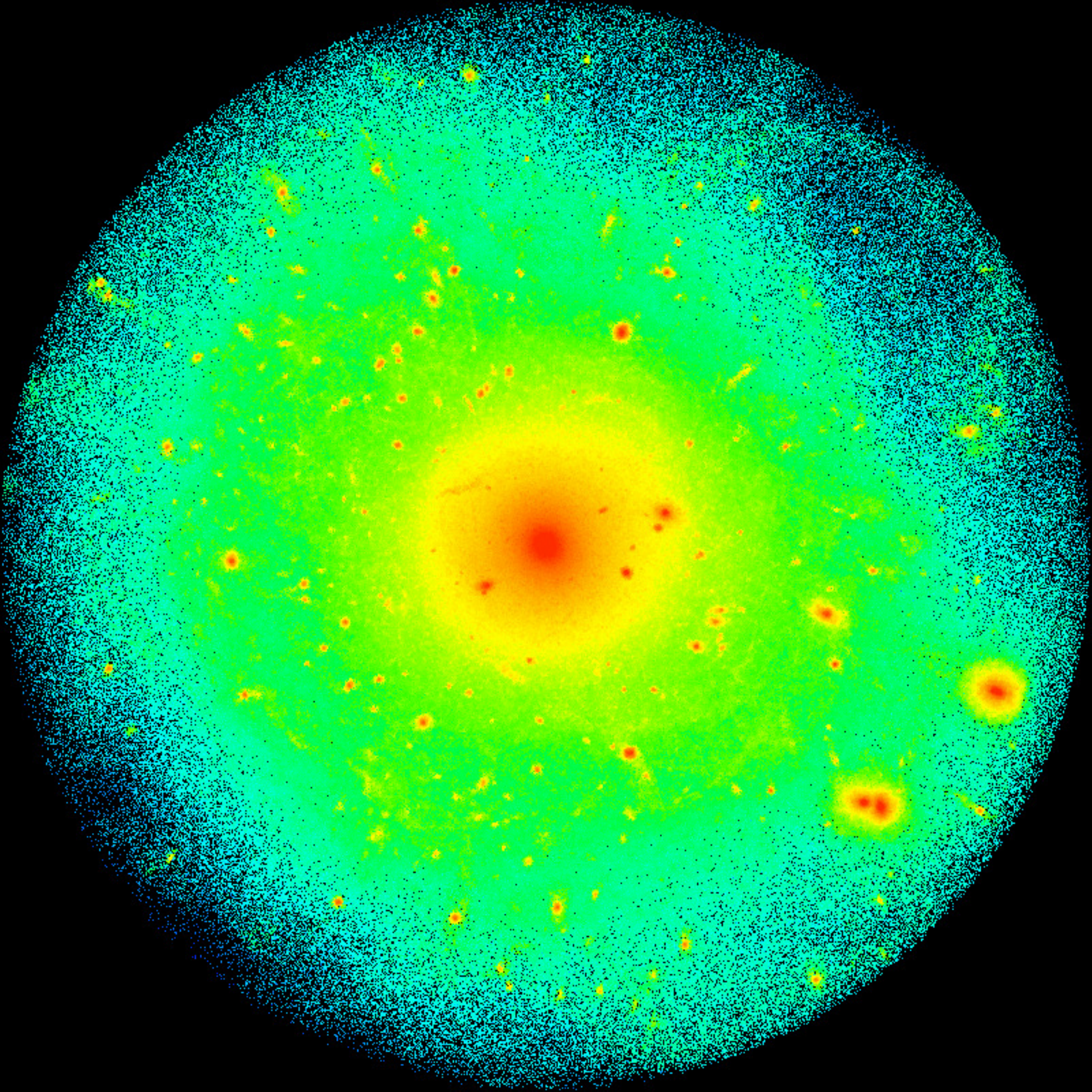}
\includegraphics[scale=0.1338]{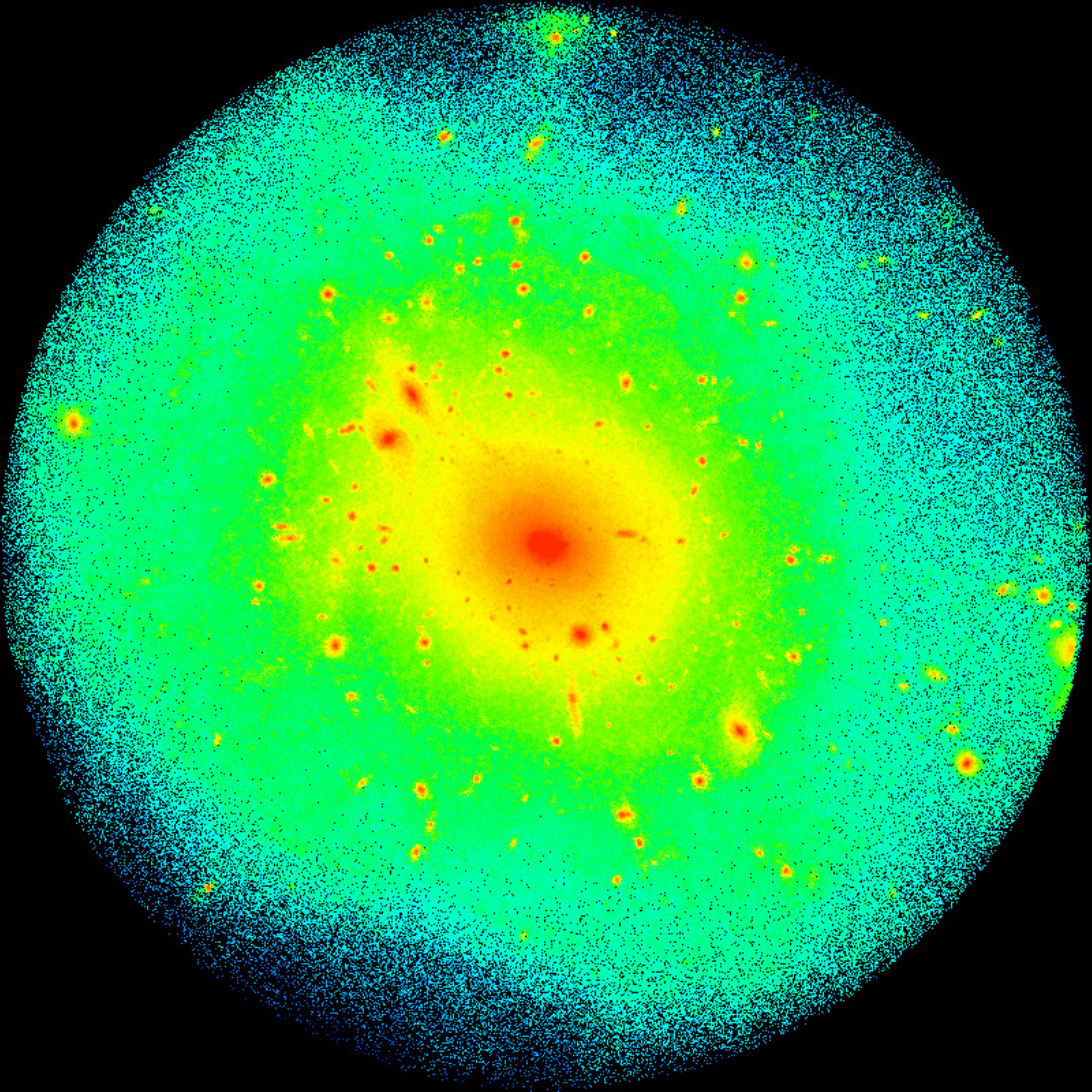}
\caption{Projected density maps of the refined Milky Way sized halo, varying the cosmology. Each box shows a sphere with radius $r_{200}$. Top row from left to right: CDM, WDM 2keV, f05-0.1keV, f20-0.3keV. Bottom row from left to right: f80-1keV, f05-0.05keV, f50-0.3keV, f20-0.1keV.}
\label{densitymaps}
\end{center}
\end{figure}

Finally, subhalos in all refinement runs are identified using 6dFoF \cite{Diemandetal2006}, a phase-space extension of the well known FoF method. The minimum number of particles in a subgroup is set to $N_{\rm min}=64$. The subhalo properties such as mass and maximum circular velocity are computed with a profiling routine that takes the phase-space coordinates of 6dFoF as input. Details can be found in \cite[][]{Zempetal2012}. An overview of all six mixed dark matter simulations and the corresponding pure CDM and pure WDM runs can be found in table \ref{simulationoverview}; the projected density maps are shown in figure \ref{densitymaps}.

\subsubsection*{Thermal Velocities and Artificial Structure}
Due to a decoupling when still relativistic, WDM particles are expected to retain a thermal motion that is assumed to influence the internal structure of dark matter haloes \cite[e.g][and references therein]{villa11,maccio2012}. We therefore added a thermal velocity component to all WDM particles on top of the Zeldovich velocities at the starting redshift $z_{\rm IC}=99$, by following the method presented in refs. \cite{maccio2012,Anderhaldenetal2012}. However, evidence suggests that large streaming velocities give rise to artificial small scale power \cite{Colin2008,DalalAnderhalden2013}. In order to quantify the influence of thermal velocities on our results, we rerun one of the most extreme simulations, i.e. f50-0.3keV, without additional thermal motion. The dashed lines in figures \ref{massfunction} and \ref{vcircprofiles} indicate that additional streaming velocities corresponding to the cosmologies at hand only impact scales below our resolution limit and do not change any of the results presented in this work.

Moreover, simulating cosmologies with a cutoff in the power spectrum leads to spurious halo formation at small scales \cite{WangWhite}. These artifacts usually form in filaments, but they may fall into larger halos and artificially enhance the substructure abundance. For the WDM run we can easily estimate the typical scale where spurious structure formation starts to occur. An extrapolation from measurements of Schneider et al. (2012) \cite{Schneideretal2012} leads to approximate mass constraint of $M \sim 10^8$ M$_{\odot}$, above which no artificial halos are expected to form. Here we have taken into account the fact that $M\propto L/N^{1/3}$, where $L$ is the box size and $N$ is the particle number of the simulation. This estimated constraint roughly corresponds to our resolution limit, we therefore do not expect a significant amount of spurious structures that would change our results.
For the MDM runs we do not expect spurious halos to form because there is a step and no cutoff in the linear power spectrum (see figure \ref{gridpower}). This means that both dark matter fluids have physical power on all scales, which triggers physical collapse and the formation of real structure.

\section{Subhalos in C+WDM}\label{results}
Dark matter subhalos have implications for both dark matter detection and satellite galaxy properties \cite[see e.g.][for recent reviews]{DiemandMoore2011,Kuhlenetal2012}.
Here we present the abundance, spatial distribution and mass profiles of subhalos. Comparing all of these quantities with observations of satellite galaxies provides a solid test of the underlying cosmology.

Some models of galaxy formation connect the most massive subhalos at redshift $z=0$ with the most luminous dwarf galaxies observed in the Milky Way \cite[e.g.][]{Stoehretal2002,Penarubbiaetal2008}. However, since subhalos undergo tidal stripping that can lead to a substantial mass loss after infall, more recent studies suggest that the most luminous Milky Way dwarfs should actually be linked to the most massive subhalos at the time of accretion \cite[e.g.][]{Bullocketal2000,Kravtsovetal2004,Bushaetal2010}.


Instead of halo and subhalos masses we use the maximum circular velocity, $v_{\rm max}$\footnote{The maximum circular velocity is defined at that radius $r_{\rm max}$ at which $v_{\rm c} = \sqrt{GM(<r)/r}$ gets maximised.}, 
which is a well defined proxy for halo and subhalo size and easy to determine accurately from cosmological N-body simulations (both at $z=0$ and at infall).
Extracting the maximum circular velocity at the time of accretion ($v_{\rm peak} \equiv v_{\rm max}(z_{\rm infall})$) for each halo is computationally more expensive than
just taking the corresponding value at redshift $z=0$ (i.e. $v_{\rm max}(z=0)$), since it requires the assembly of a merger tree. 

In appendix \ref{method} we present a simple method how to compute an estimate for $v_{\rm peak} = v_{\rm max}(z_{\rm infall})$ of a subhalo, based on its present-day properties alone ($v_{\rm max}(z=0)$ and binding energy).
Although afflicted with some scatter, it turns out to be a useful estimator for the maximum circular velocity at infall and provides quick method for a comparison with observational data.
For the following analysis, we will make use of two selection criteria, $v_{\rm max}$ and $v_{\rm peak}$, whenever comparing with observational data.

\subsection{Abundance}
\begin{figure}
\begin{center}
\includegraphics[scale=0.38]{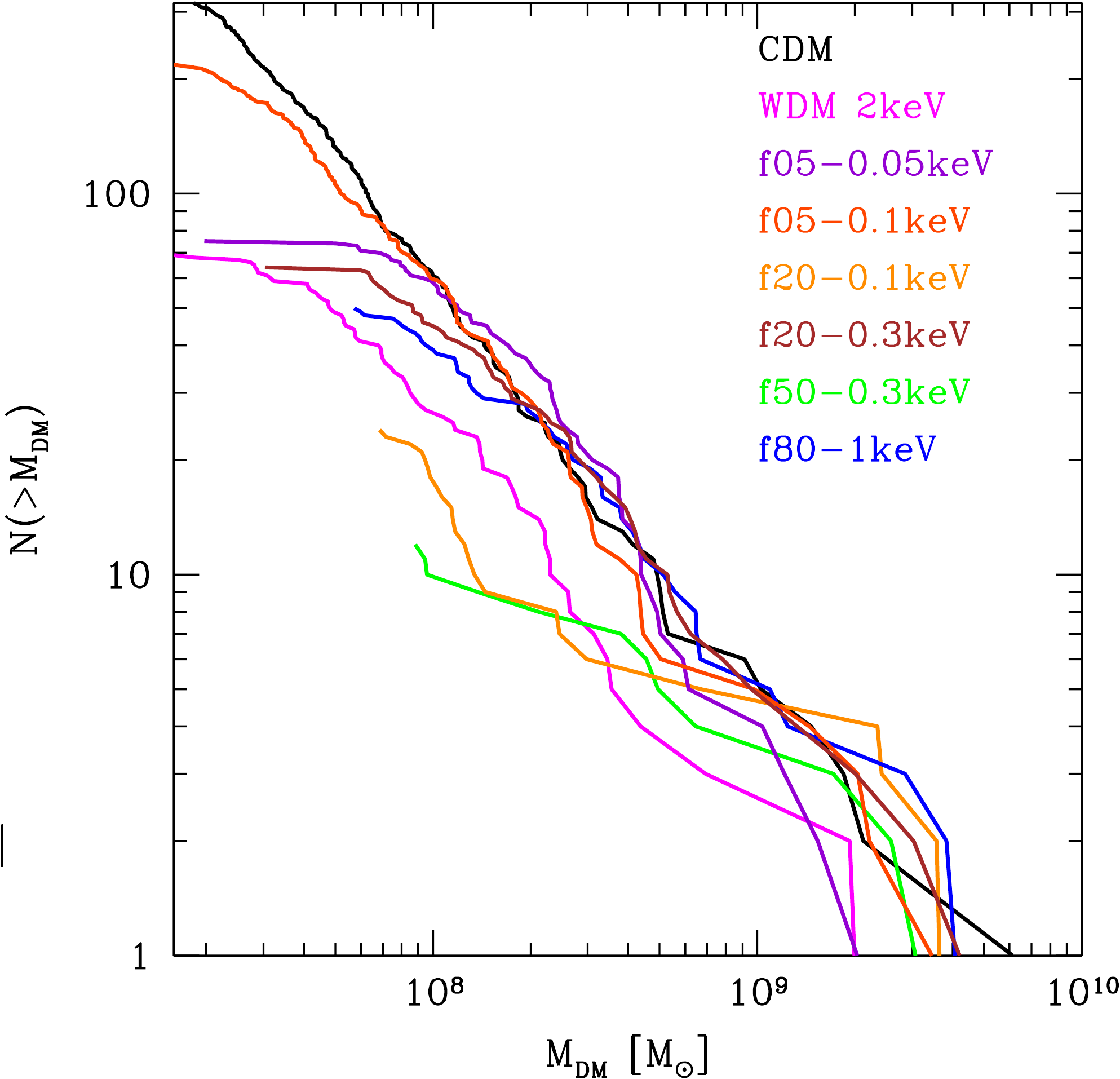}
\includegraphics[scale=0.38]{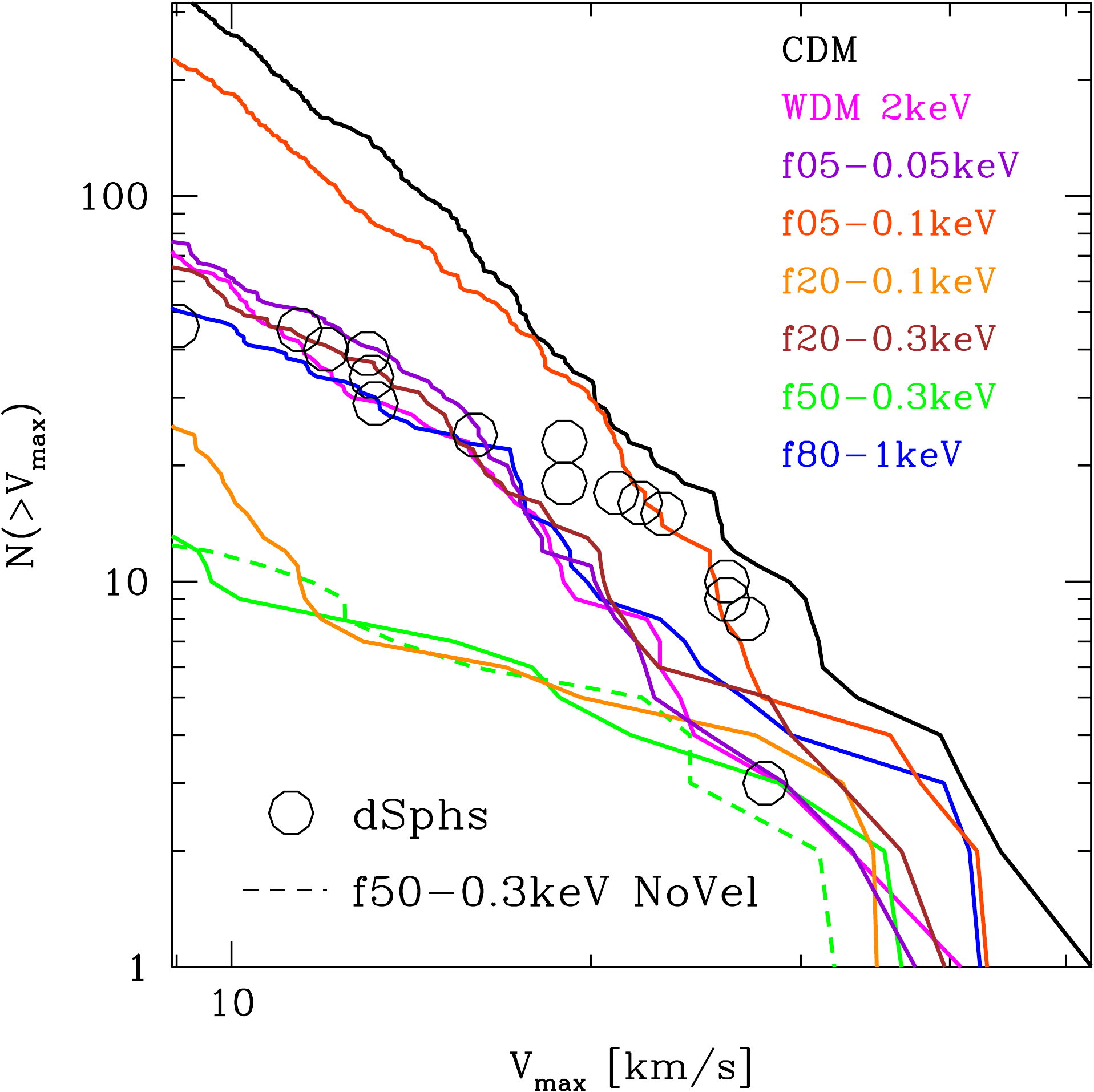}
\caption{Subhalo abundances in all cosmologies, represented as a function of subhalo tidal mass (left panel) and as a function of the maximum circular velocity $v_{\rm max}$ at redshift $z=0$ (right panel). Estimates for the $v_{\rm max}$ of the host subhalos of the Milky Way satellite galaxies are shown for comparison (open circles, right panel) \cite[][]{Wolfetal2010,Polisensky2011,Boylan-Kolchinetal2011}. Note that converting from the observed stellar kinematics to the much larger scale of $v_{\rm max}$ introduces large uncertainties and relies on CDM subhalo scaling relations \cite[][]{Strigarietal2007}. In addition, the right panel also shows the subhalo velocity function in the f50-0.3keV model without thermal velocities (green dashed line). No striking differences are evident.}
\label{massfunction}
\end{center}
\end{figure}
In the past years, the SDSS survey has lead to the discovery of more than 10 new very faint satellites \cite[e.g.][]{Belokurovetal2007,Koposovetal2008}. This has allowed the possibility to extend (or circular velocity) the comparison between simulations and observations to lower masses (or circular velocity) \cite[][]{Strigarietal2007}.
The left panel of figure \ref{massfunction} shows the subhalo mass function, the right panel shows the cumulative velocity function for all subhalos with a value of $v_{\rm max}>9$ km s$^{-1}$ (which is approximately the resolution limit of the CDM reference simulation), as well as a sample of observed Milky Way satellites (black open circles). When representing the subhalo abundance as a function of tidal mass, most of the non-standard models closely follow the CDM case. On the other hand, plotting the subhalo abundance as a function of $v_{\rm max}$, the differences get larger since $v_{\rm max}$ is sensitive to the concentration of the subhalo.
The observation sample consists of both classical and SDSS satellites, though for the latter, we included a completeness correction since the SDSS only covers about 28 per cent of the sky. Whereas the CDM cumulative velocity function is well approximated by a power law \cite[e.g.][]{Diemandetal2008,Klypinetal2011},
\begin{equation}
N(> v_{\rm max}) \sim (v_{\rm max} / v_{\rm max, host})^{-3},
\end{equation}
all cosmologies containing some fraction of warm dark matter exhibit a reduced total abundance of substructure (and therefore a shallower velocity function) due to the lack of power on small scales. As shown in \cite{Anderhaldenetal2012}, a generic mixed dark matter cosmology containing a thermal WDM candidate can be brought into a hierarchy according to its effective free-streaming length, i.e.
\begin{equation}
\label{effectivefs}
\lambda_{\rm fs}^{\rm eff} \propto \Big(\frac{f_{\rm W}}{m_{\rm WDM}} \Big)^{4/3}.
\end{equation}
Hence, the C+WDM models f20 - 0.1keV and f50 - 0.3keV are the most extreme in our sample. This is not only apparent in the linear power spectrum (right panel of figure \ref{gridpower}) but also in the subhalo velocity functions  and mass functions (orange and green solid lines): only a few subhalos exist in galactic halos for these two models.

For comparison, Figure \ref{massfunction} also shows the estimated $v_{\rm max}$ values for the observed Milky Way satellite galaxies. Since the host subhalo $r_{\rm max}$ lies well beyond the satellites, these estimates are very uncertain and it is more accurate to compare simulated subhalo mass profiles with observations at much smaller radii \cite[][]{Strigarietal2007}, especially the half light radii \cite[][]{Wolfetal2010}, which we will do in section \ref{massprofiles}.
Using the approximate satellite galaxy $v_{\rm max}$ values for now, we find that pure 2keV WDM scenario as well as the mixed dark matter models f80 - 1keV, f20 - 0.3keV and f05 - 0.05keV reproduce the observed abundance quite well. In the pure CDM and f05 - 0.1keV case, the classical problem of missing satellites is clearly visible and some mechanism to suppress star formation in some fraction of these subhalos is required to match the velocity function estimated from the observations. It is possible that with increasing sensitivity of observational methods over the next several years, many more faint and ultra-faints dwarfs will be detected, which might continuously close the gap between the number of observed dwarf galaxies and simulated CDM subhalos \cite[e.g.][]{Tollerud2010}. For a very solid, conservative constraint on the underlying cosmology we therefore just demand that the number of currently know satellites be comparable or larger than the number of subhalos in the corresponding size range. This lower bound on the required subhalo abundance clearly rules out the two extreme mixed models f20 - 0.1keV and f50 - 0.3keV.


\begin{figure}
\begin{center}
\includegraphics[scale=0.38]{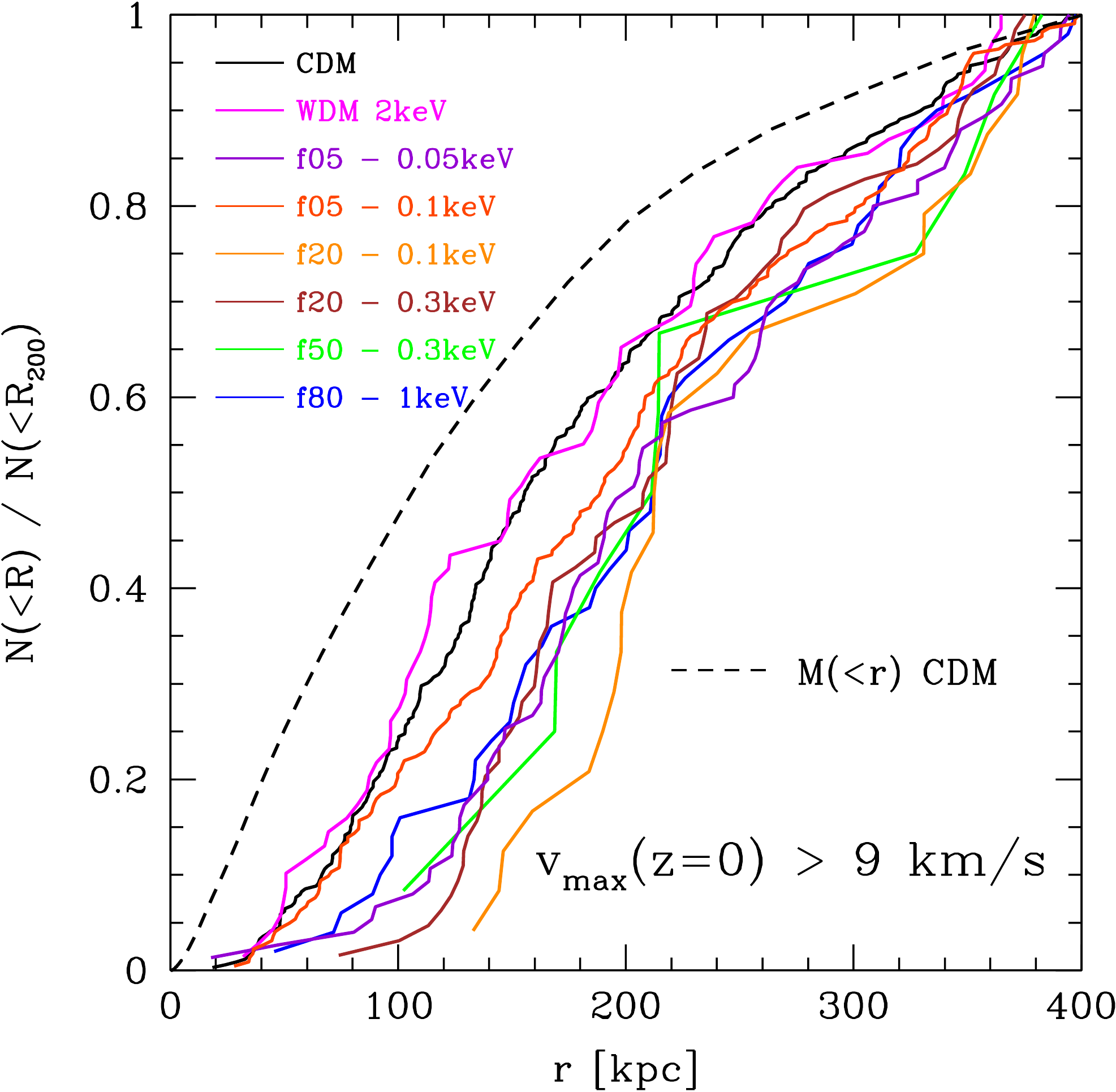}
\includegraphics[scale=0.38]{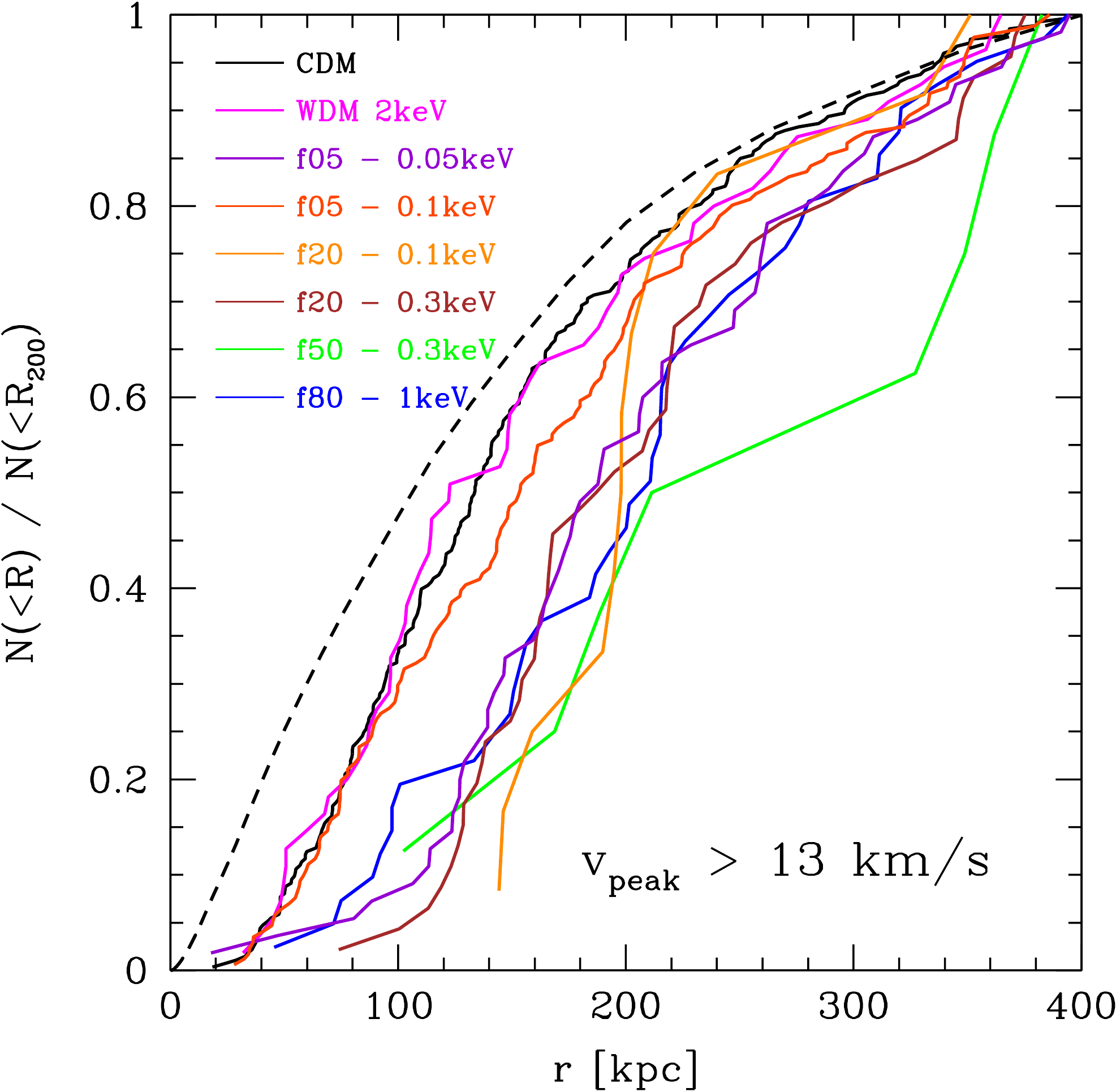}
\newline
\includegraphics[scale=0.38]{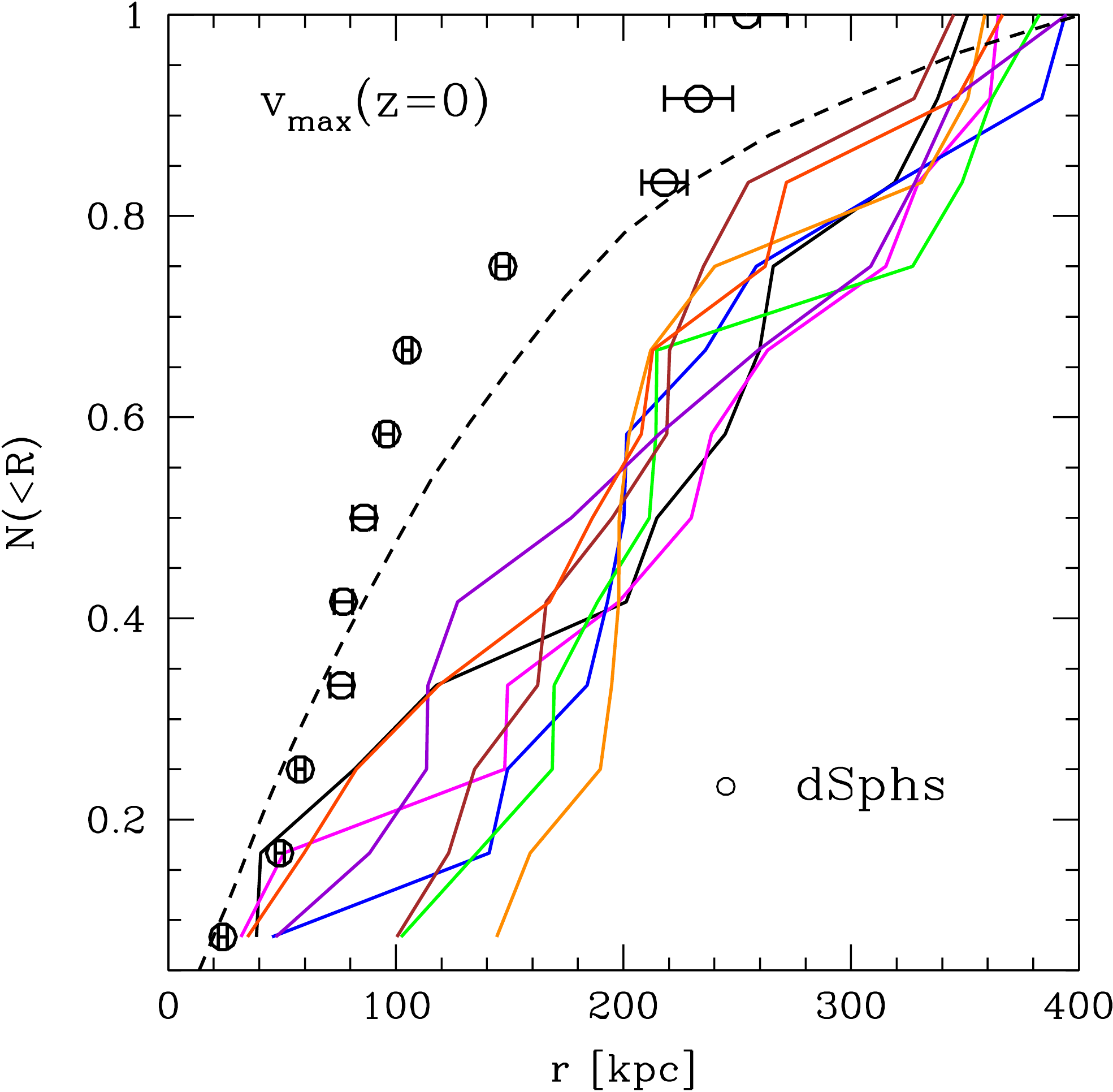}
\includegraphics[scale=0.38]{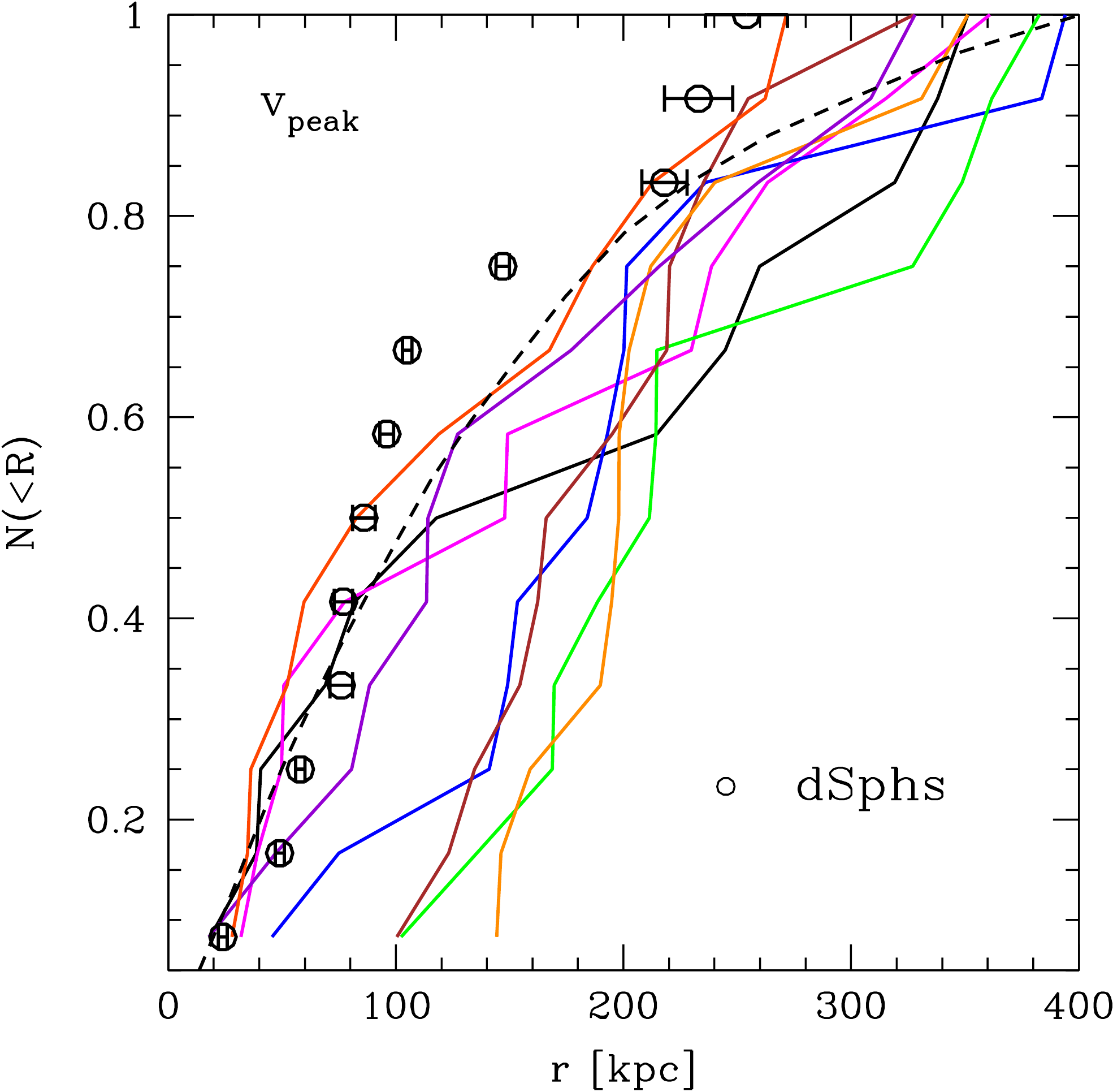}
\caption{Top: Normalised radial distribution of all subhaloes within $R_{200}$ that have that have $v_{\rm max} > 9$ km/s (left) and $v_{\rm peak} > $13 km/s (right) respectively. Bottom: Normalised radial distribution of the 12 subhalos with the largest values of the maximum circular velocity, selected after $v_{\rm max}$ (left) and $v_{\rm peak}$ (right). Moreover, the normalised mass distribution of the main CDM halo is shown in both panels (black dashed line). The open circles refer the 12 most luminous satellites observed. Horizontal error bars refer to observational uncertainties. Starting with the smallest distance those are: Sagittarius, LMC, SMC, Ursa Minor, Draco, Sculptor, Sextans, Carina, Fornax, Leo II, Canes Venatici I, Leo I (data taken from \cite[][]{Wolfetal2010,Polisensky2011,Boylan-Kolchinetal2011}).}
\label{radialdistribution}
\end{center}
\end{figure}
\subsection{Radial Distribution}
The radial distribution of satellite galaxies provides another important observational constraint: it needs to be matched by any realistic set of subhalos selected from a cosmological simulation in order to be possible hosts of the satellite galaxies. The free streaming effect in warm dark matter models does not only suppress the subhalo abundance, but they are also found to have later formation times and therefore lower concentrations when compared with CDM \cite[e.g.][]{Maccioetal2010,Lovelletal2012}. Lower concentration subhalos are more affected by tidal mass loss and less likely to survive in the inner halo, where some of the satellite galaxies are observed.

Figure \ref{radialdistribution} shows the normalised subhalo radial distribution of all six mixed dark matter models as well as the two pure CDM and WDM runs. In the two panels on top, all subhalos with maximum circular velocity $v_{\rm max} > 9$ km/s (left) and $v_{\rm peak} > $ 13 km/s (right) are used. The bottom panels only shows the 12 largest objects according to each of the two selection criteria used above ($v_{\rm max}$ on the left and $v_{\rm peak}$ on the right) as well as the 12 most luminous satellites observed. These consist of 11 classical dSphs (Sagittarius, LMC, SMC, Ursa Minor, Draco, Sculptor, Sextans, Carina, Fornax, Leo II, Leo I) plus Canes Venatici I with a luminosity comparable with Draco but only observed in the SDSS survey. The black dashed line refers to the normalised mass distribution of the CDM halo.

As already mentioned, there is a theoretical motivation to believe that $v_{\rm peak}$ correlates better with galaxy luminosity than $v_{\rm max}$. We will therefore mainly focus on the bottom right panel of figure \ref{radialdistribution}. For small galactocentric distances ($r \lesssim 100$ kpc), the two models containing only a small WDM fraction as well as the pure 2 keV WDM model match the observed satellite distribution remarkably well. On the other hand, in mixed cosmologies with larger free-streaming length, i.e. f20 - 0.1keV, f20 - 0.3keV and f50 - 0.3keV, none of the 12 most massive subhalos reside on orbits close ($r \lesssim 100$ kpc) to the galactic center.

\subsection{Mass Profiles}
\label{massprofiles}
\begin{figure}
\includegraphics[scale=0.25]{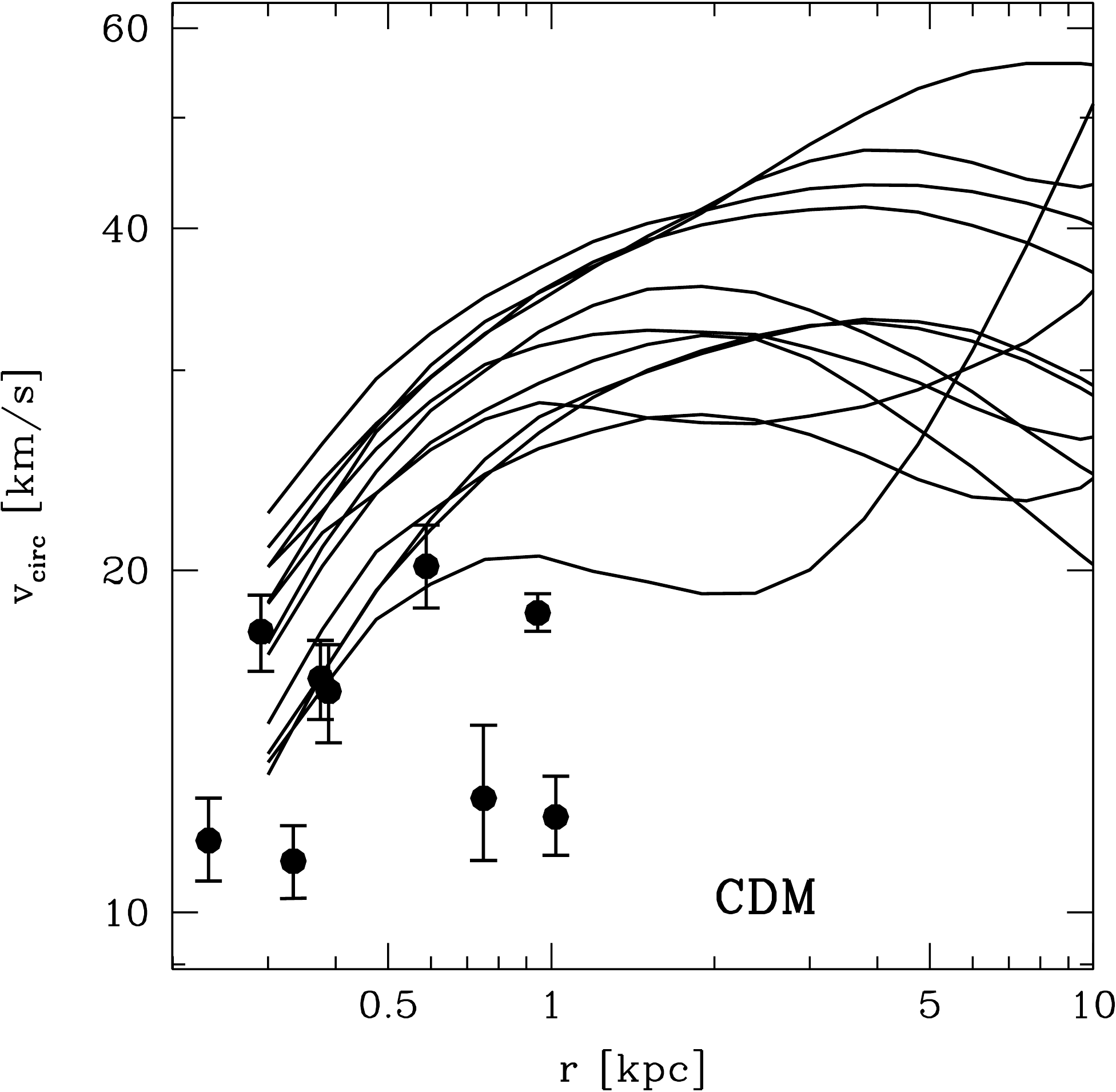}
\includegraphics[scale=0.25]{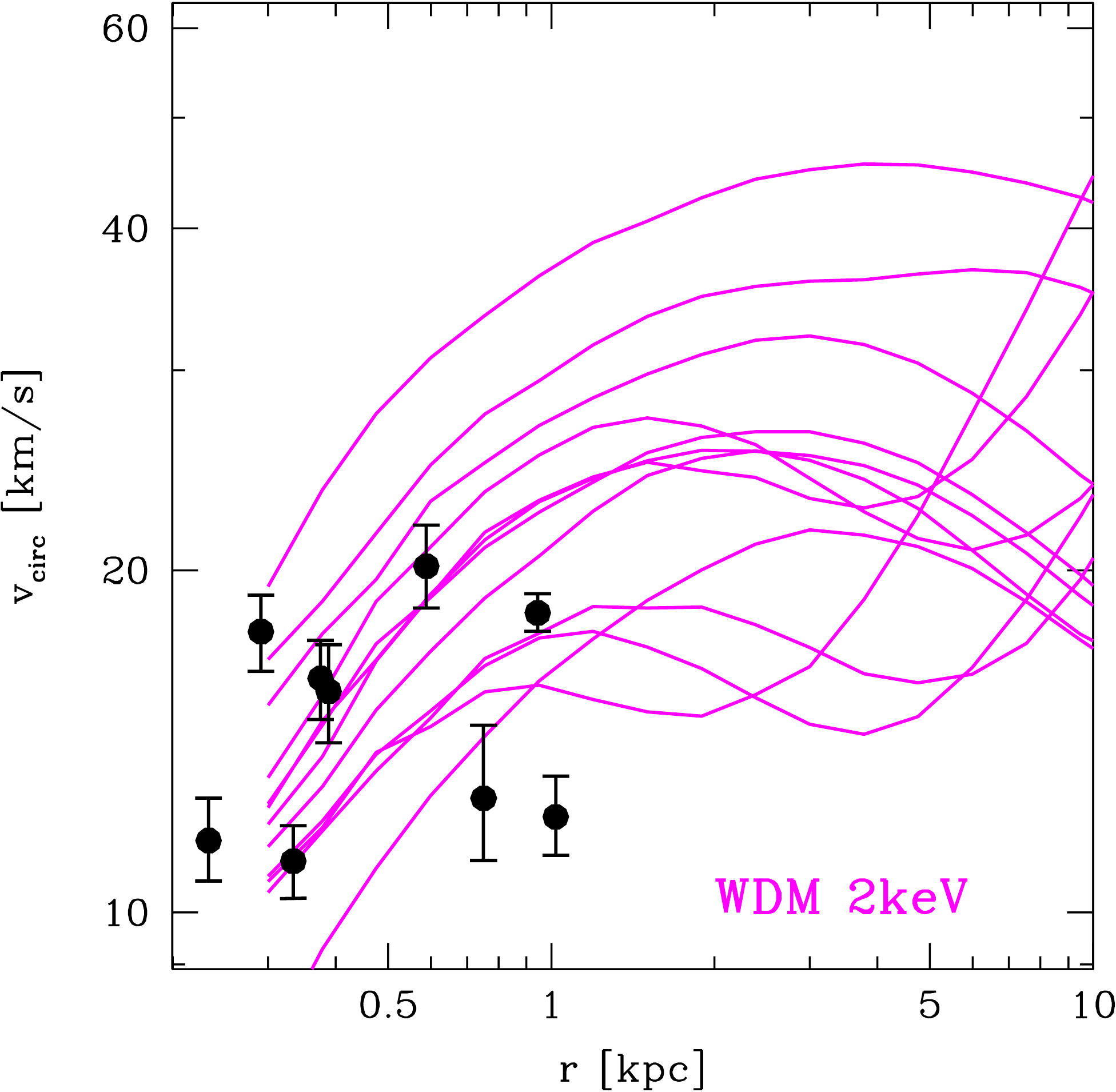}
\includegraphics[scale=0.25]{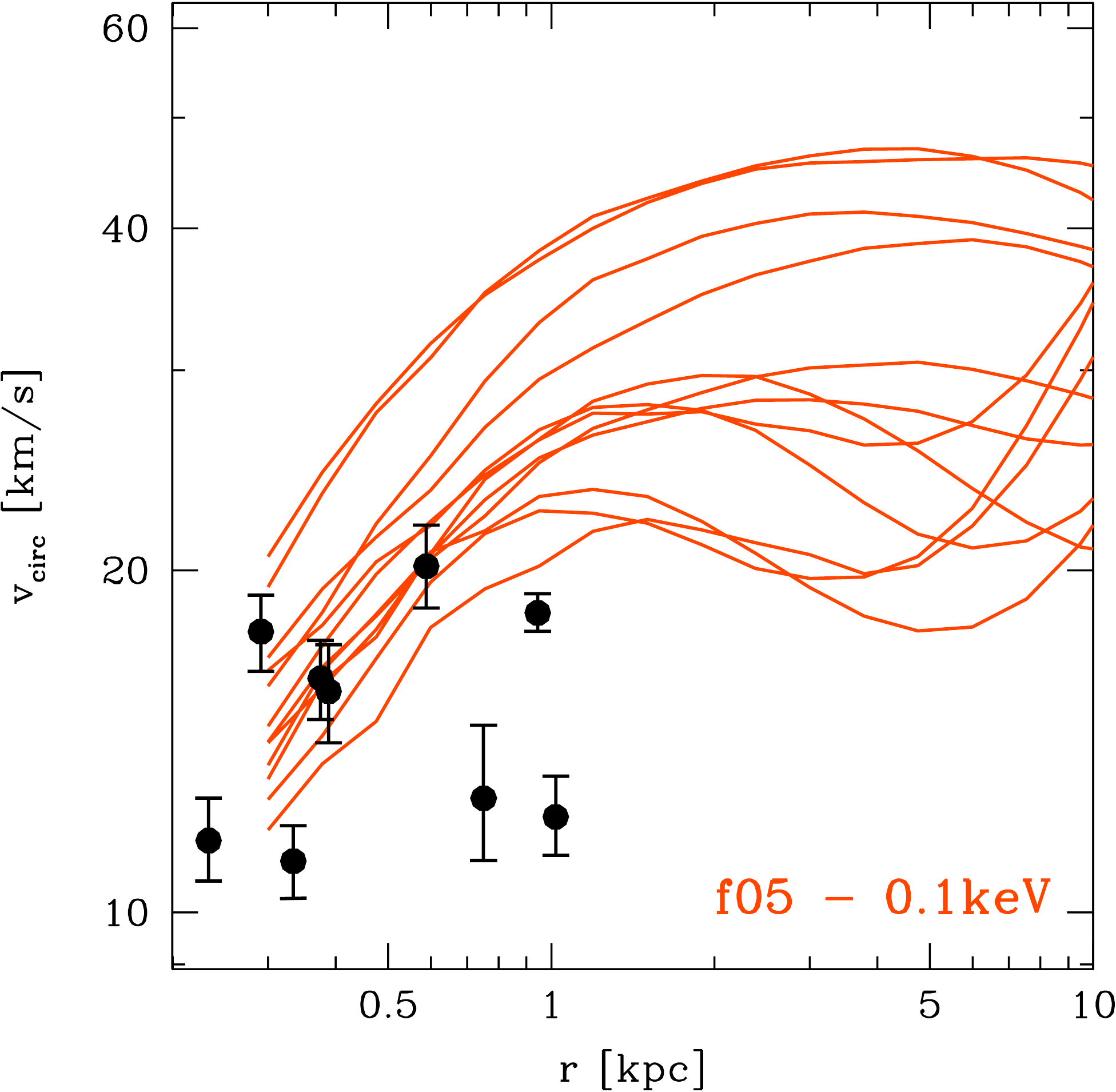}
\newline
\includegraphics[scale=0.25]{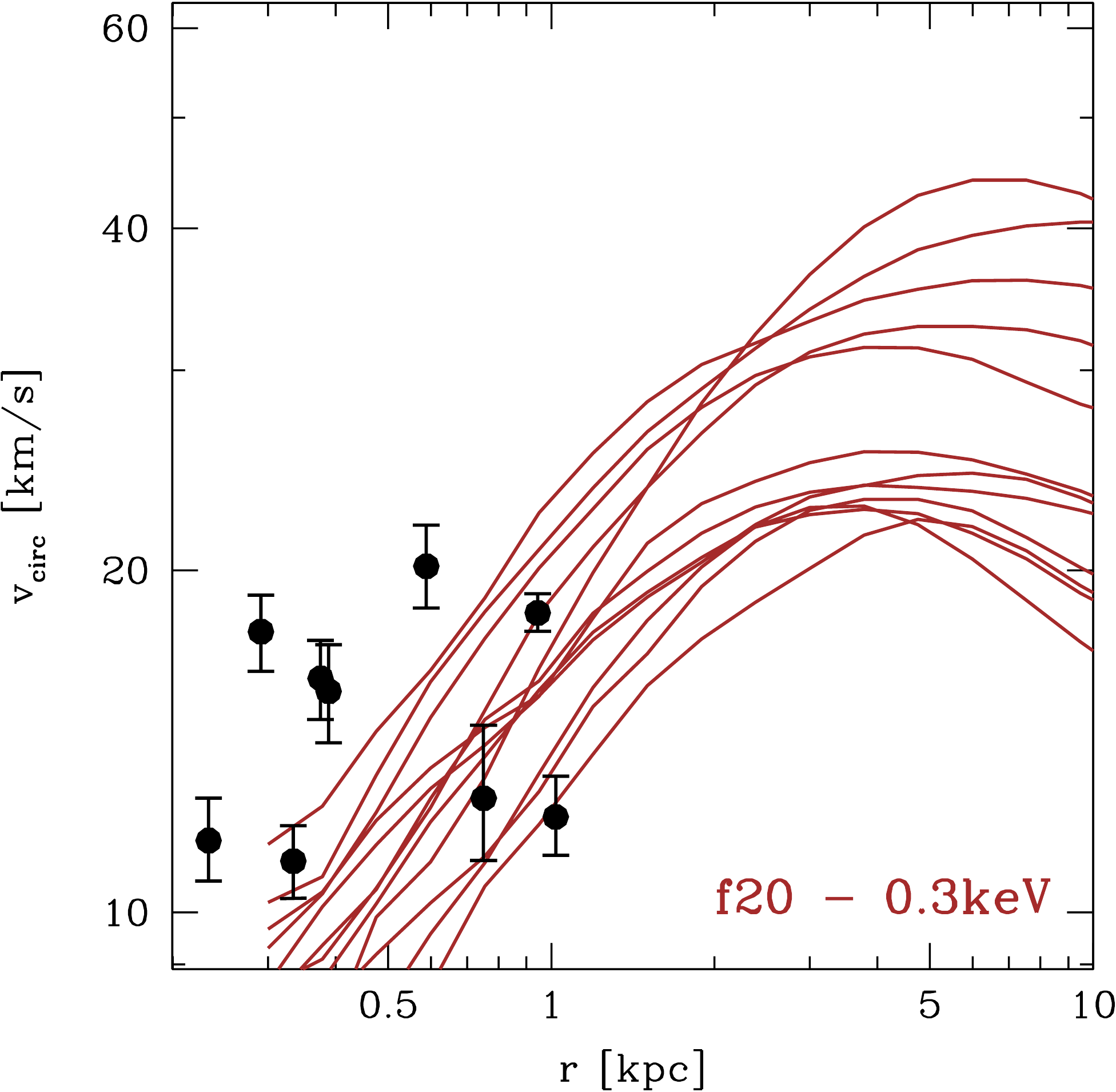}
\includegraphics[scale=0.25]{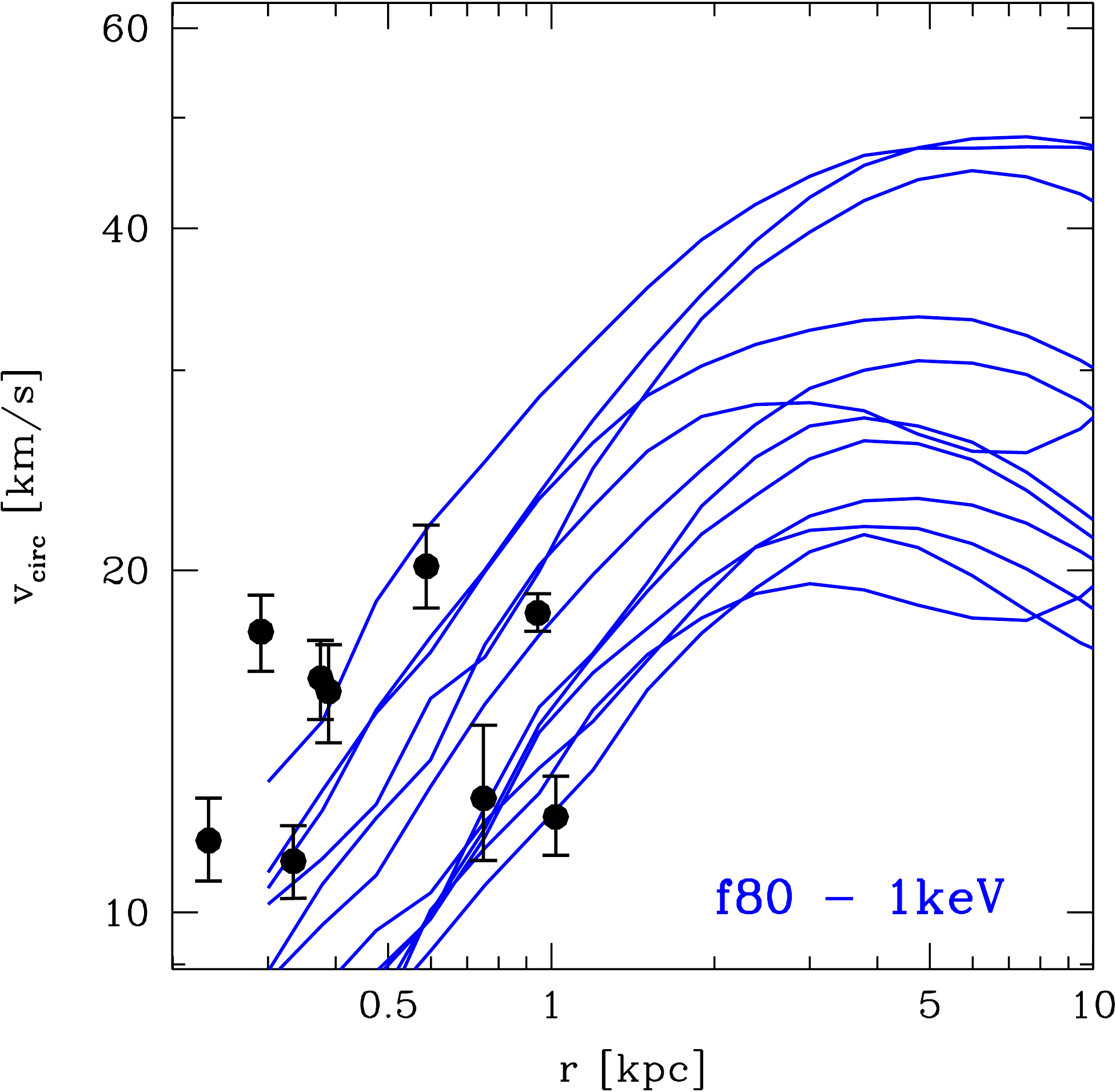}
\includegraphics[scale=0.25]{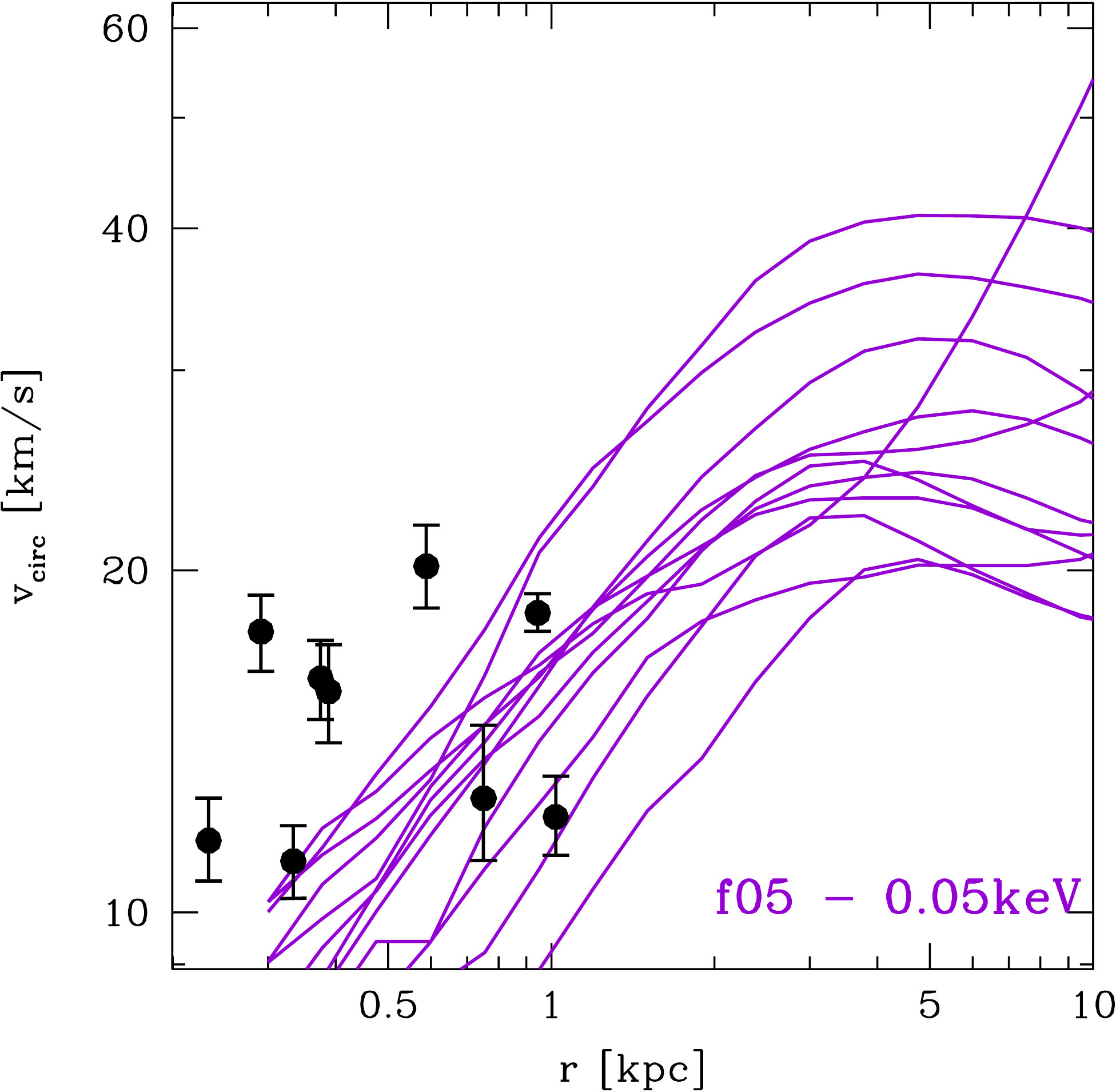}
\newline
\includegraphics[scale=0.25]{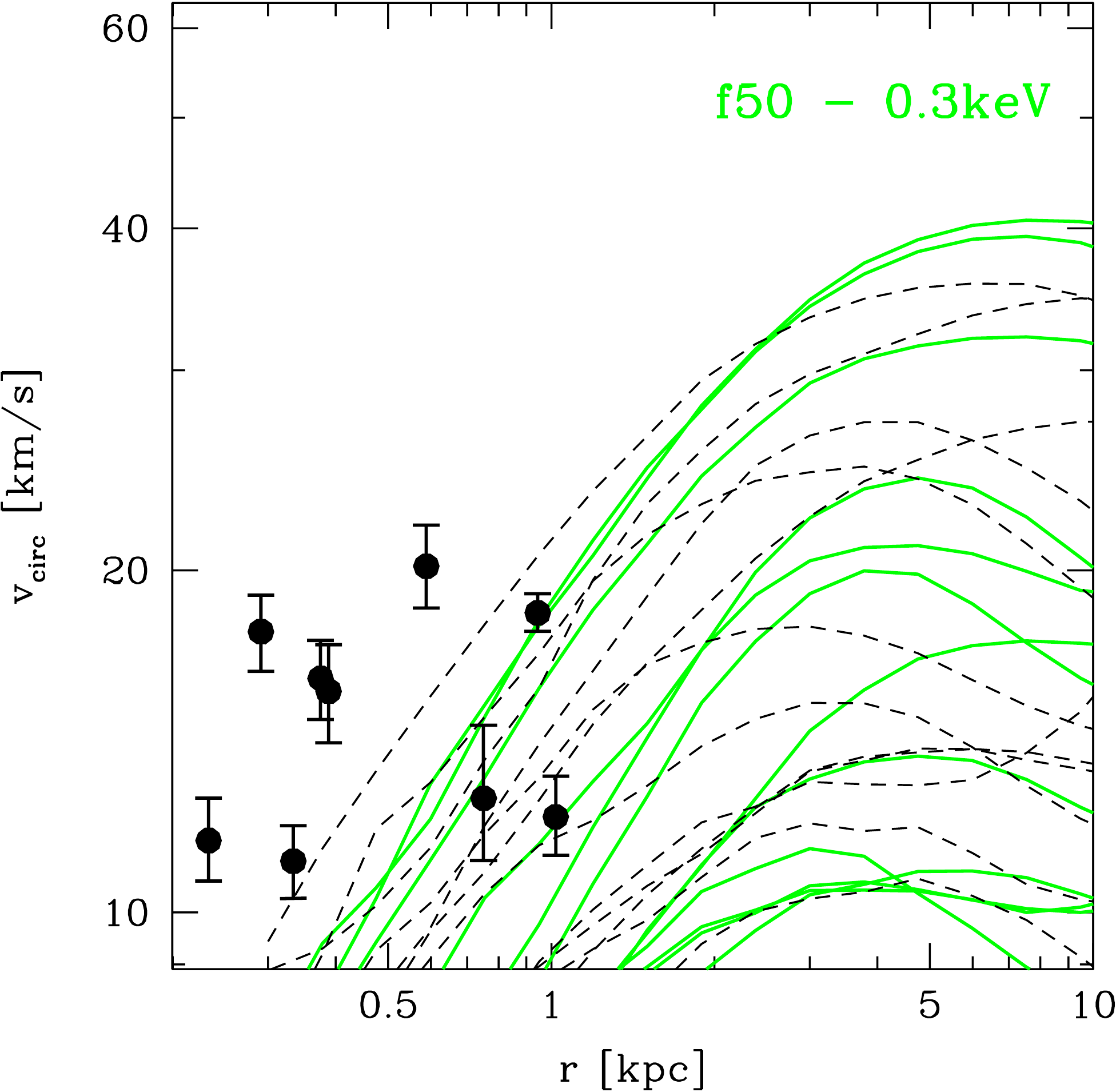}
\includegraphics[scale=0.25]{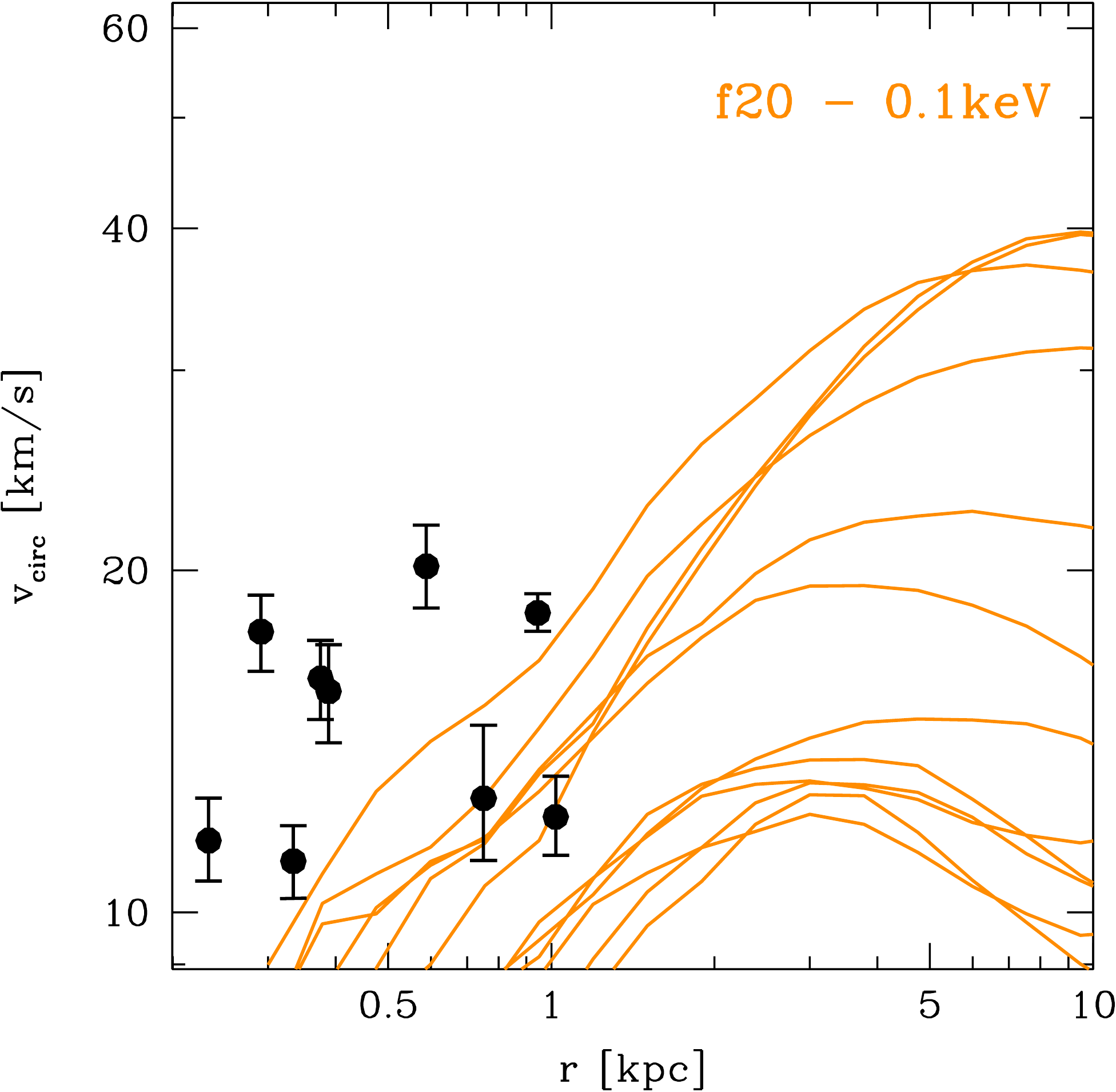}
\caption{Circular velocity curves at redshift $z=0$ for the 12 subhaloes that had the highest value of $v_{\rm peak}$. Measured $v_{\rm circ}(r_{1/2})$ values for the brightest dwarf galaxies of the Milky Way are plotted as data points with error bars. In contrast to Fig. \ref{radialdistribution}, LMC, SMC and Sagittarius are removed from the sample. In all panels, the total mass distribution around the subhalo centers is plotted. The upturn in some of the profiles is due to the mean halo background density. The additional black dashed lines in the lower left panel show the f50-0.3 keV simulation without thermal velocities (see Section 2).}
\label{vcircprofiles}
\end{figure}

It was recently shown that internal structure and abundance of the bright Milky Way satellites is not consistent with populating the largest CDM subhalos in main halos of $M_{\rm vir} \gtrsim1.7 - 2.0 \times 10^{12}$ M$_{\odot}$ \cite[][]{Boylan-Kolchinetal2011,Boylan-Kolchinetal2012,Rashkovetal2012}. To match the observations some of the largest, densest CDM subhalos would have to fail to form a luminous dwarf galaxy, which seems rather unlikely. This issue has been
quite accurately and very timely referred to as the "too big to fail" problem \cite[][]{Boylan-Kolchinetal2011}.
However, it turns out that most but not all of the halos in this mass range do contain too many large subhalos \cite[e.g.][]{PurcellZentner2012}, which raises the question why the Milky Way halo should be special. Since the discrepancy in the Via Lactea II simulation is only 
about a factor of two in mass \cite[][]{Rashkovetal2012} rescaling this halo to a smaller virial mass of ($M \sim 0.8 \times 10^{12}$ M$_{\odot}$) would resolve the discrepancy (see also \cite[e.g.][]{LiWhite,Watkinsetal2010,Bushaetal2011}). However, even if the presence of two massive objects as the Magellanic clouds \cite{Boylan-Kolchinetal2010} disfavour such a low halo mass for the Milky Way, the exact virial mass of the galactic halo is still an ongoing debate and ranges from $M \sim 0.6 \times 10^{12}$ M$_{\odot}$ up to $M \sim 1.9 \times 10^{12}$ M$_{\odot}$ \cite[e.g.][]{Xueetal2008,Gnedinetal2010,Watkinsetal2010,Kafleetal2012}.

Other ways to reconcile CDM with these constraints are discussed in the literature: For example baryonic feedback processes or highly stochastic galaxy formation in the most dense subhalos but it is currently unclear how plausible they really are \cite[e.g.][and references therein]{Boylan-Kolchinetal2012}.

Non-standard cosmologies like self-interacting, warm or mixed dark matter might provide a solution to this issue and they have recently attracted a lot of interest \cite[e.g.][]{Lovelletal2012,Anderhaldenetal2012,Vogelsbergeretal2012}. Lovell et al. (2012) \cite{Lovelletal2012} have re-run the Aquarius A halo in a thermal 1.4 keV warm dark matter cosmology. Indeed, they find that subhalos are not only less numerous but also less concentrated (due to their later formation time) than their CDM counterparts. Therefore, a WDM cosmology with a thermal dark matter candidate with a mass of roughly 2 keV does not suffer from the "too big to fail" problem.

Figure \ref{vcircprofiles} shows the circular velocity profiles at redshift $z=0$ for the 12 subhalos with the largest value of $v_{\rm peak}$ for all cosmologies under consideration. As mentioned above, it is evident that CDM predicts several subhalos which are too massive in order to be possible hosts of the most luminous satellites. Due to the reduced power at small scales, structure formation in all other cosmologies is delayed, leading to slightly smaller halo and subhalo concentrations. When going towards models with increasing effective free streaming length (from left to right and from top to bottom), the central masses of the subhalos are substantially lowered. While the pure 2 keV WDM, f05 - 0.05keV and f80 - 1keV models match the observed satellite galaxies quite well, the remaining cosmologies, especially the two outlying models f50 - 0.3keV and f20 - 0.1keV, fall into the other extreme: they do not produce enough massive subhalos in order to host the observed satellite galaxies.

\section{Discussion \& Conclusions}\label{discussion}

So far, the best observational constraints on the non-linear matter power spectrum, and therefore on the nature of dark matter are given by the Lyman alpha forest (Ly-$\alpha$) \cite[e.g.][]{Vieletal2005}. Assuming a pure warm dark matter cosmology, Ly-$\alpha$ analysis provide a lower bound on the dark matter particle of roughly the order of one keV (strongly dependent on the production mechanism). In this paper we have demonstrated how satellite galaxies of the Milky Way with masses $10^{\sim (7-9)}$ M$_{\odot}$ allow to test dark matter on even smaller scales than the Ly-$\alpha$ constraints, i.e. we can probe (and rule out) free-streaming scales, which lie significantly below the Ly-$\alpha$ constraints.
Therefore, a detailed comparison between observations and subhalos in a numerical simulation is a further powerful test to constrain the nature of dark matter.

We have performed a set of of high resolution N-body simulations of a Milky Way sized dark matter halo in a variety of cosmological models: CDM, WDM with a particle mass of 2 keV as well as six different mixed dark matter models (C+WDM) varying the fraction and the mass of the warm component (see left panel of Fig. \ref{gridpower} for an overview). The most massive subhalos in each simulation had then to pass three tests in order to be considered as a viable cosmological model. They should simultaneously reproduce the abundance of satellites, their radial distribution and their mass profile. Our findings can be summarised as follows:
\begin{itemize}
  \item Comparing the subhalo {\it abundance} with observations only provides a lower bound on a given cosmology in terms of its free streaming length. The number of subhalos in the models f20 - 0.1keV and f50 - 0.3keV lies clearly below the observed satellite galaxy abundance and therefore these models are ruled out. CDM and f05 - 0.1keV offer more than enough subhalos and could match the observational counts if some strong feedback mechanism suppresses galaxy formation in small halos, or in other words, both these models may suffer from the well known "missing satellites problem". The three remaining C+WDM models (f05 - 0.05keV, f20 - 0.3keV \& f80 - 1keV) as well as the pure 2keV WDM model are in good agreement in the range $v_{\rm max}\lesssim 20$ km/s, i.e. they would produce a reasonable number of satellite galaxies, if each subhalo contains a galaxy.
  \item Considering also the  {\it radial distribution} of subhalos, two additional C+WDM cosmologies are at the borderline of being ruled out. The largest subhalos (selected at infall time) of the models f20 - 0.3keV and f80 - 1keV reside almost entirely in the outer regions of the main halo, while the satellite galaxies tend to follow the main halo mass distribution. Also note that subhalo samples selected by their present (z=0) size are ruled out in all cosmologies (including CDM), because such samples
  are far too extended to match the observations.
  \item Incorporating also the {\it mass profiles} into the big picture, it seems that only the pure WDM model with a 2keV particle is able to match the all observations. The CDM halo and the f05 - 0.1keV halo contain several subhalos that are clearly too dense in order to host the brightest satellite galaxies, i.e. they both suffer from the 'too big to fail' problem,  but for the f05 - 0.1keV the discrepancy is much smaller. Model f80 - 1keV on the other hand, does quite well in the mass profile test, but it predicts a too extended spatial distribution of satellite galaxies.
\end{itemize}
We can conclude that our detailed comparison between observations and subhalos rules out a range of mixed C+WDM models (f20 - 0.3keV, f80 - 1keV and f05 - 0.05keV), which are not excluded by the Ly-$\alpha$ constraints. The most stringent test comes from the subhalo mass profiles (figure \ref{vcircprofiles}), which rules out four models because their subhalos are not dense enough; and it challenges CDM, which has several subhalos which are too dense. One model (f80 - 1keV) gives an acceptable match in this test, but fails to match the radial distribution of observed satellites. This illustrates that for the best, strongest constraints one has to take into account both the densities and the radial distributions. The two tests are not independent however: most models, which do not form dense enough subhalos (f20 - 0.3keV, f05 - 0.05keV, f50 - 0.3keV and f20 - 0.1keV), also do not predict enough satellite galaxies in the inner halo. That is not surprising, since high central densities are needed for subhalos to resist tidal stripping in the inner halo \cite[e.g.][]{Diemandetal2007}. Furthermore, it is fair to mention that halo-to-halo scatter could also somewhat relax our conclusions.

Whether or not baryonic physics will alter these conclusions is still an open question. On the one hand, feedback processes might be able to reduce the dark matter densities near the centres of halos and subhalos \cite[e.g.][]{Parryetal2012}, and on the other hand, non-standard physics is presumably needed for regulating star formation in the smallest halos \cite[e.g.][]{WadepuhlSpringel}. However, dwarf galaxies are known to be the most dark matter dominated objects in the Universe and it is therefore reasonable to assume that spatial distribution and mass profiles of these objects are mainly driven by the dark component and not by baryons.

\acknowledgments  
It is a pleasure to thank Marcel Zemp and Darren Reed for useful comments in the final stages of this work. Further, we would like to thank the anonymous referee for helpful comments and suggestions. AVM acknowledges funding by Sonderforschungsbereich SFB 881 "The Milky Way System" (subproject A1) of the German Research Foundation (DFG). GB acknowledges the support of the European Research Council through the ERC Starting Grant "WIMPs Kairos". Numerical simulations were performed on the Theo cluster at the Max-Planck-Institut f\"ur Astronomie at the Rechenzentrum in Garching, the zBox3 cluster at the Institute for Theoretical Physics in Zurich and on the Rosa cluster at the CSCS in Lugano. This work was supported by the SNF.

\appendix
\section{Correlation between orbital energy and subhalo mass loss}\label{method}
As pointed out in the Introduction, most recent models of galaxy formation connect the most luminous satellites with the largest subhaloes at infall, i.e. the relevant subhalo size is the one measured at their first infall into the main halo ($z_{\rm infall}$) and
not the present $(z=0)$ size that is left over after tidal mass loss, which can be very large for some subhalos (especially for those near the host halo center). The size at infall can be quantified with
$v_{\rm peak} = v_{\rm max}(z_{\rm infall})$\footnote{For simplicity, $v_{\rm max}(z_{\rm infall})$ is denoted by $v_{\rm peak}$. This notation is motivated by the observation that most subhalos reach the peak vaule of their maximum circular velocity at
infall \cite{Diemandetal2007}.}. This requires for each subhalo in a simulation a merger tree built from a large number of simulation outputs. It has recently been shown that there is a strong correlation between the infall time of a subhalo and its binding energy at redshift $z=0$ \cite{Rochaetal2012,BoylanetalLeo1}. In the following, we present a strong correlation between the z=0 subhalo orbital energy and the ratio $v_{\rm max}(z=0) / v_{\rm peak}$. This provides a new and simple method how the required information, i.e. $v_{\rm peak}$, can be extracted from only present day subhalo data, without building a computationally expensive merger tree. 
\begin{figure}
\begin{center}
\includegraphics[scale=0.5]{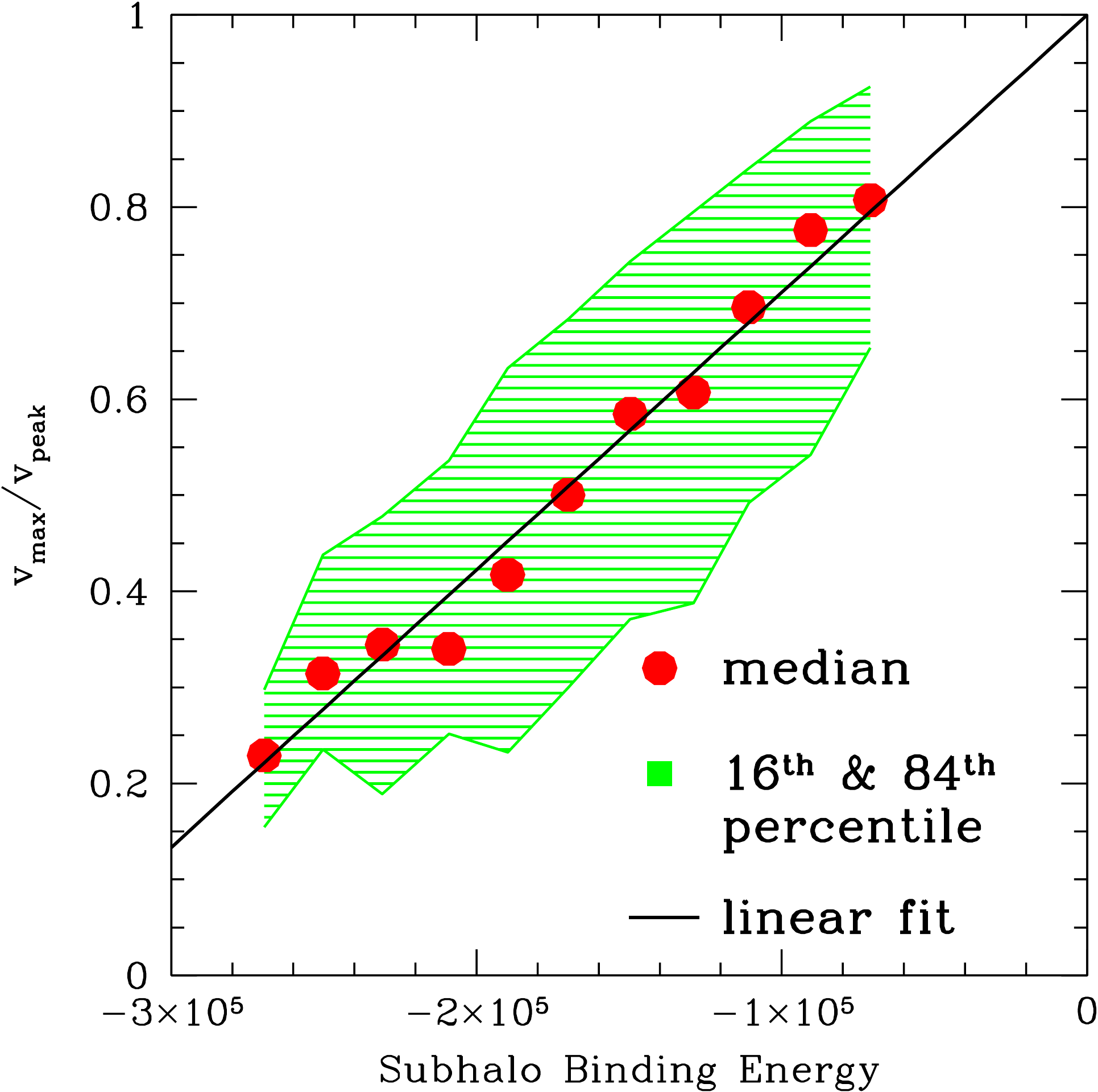}
\caption{Relation between the orbital energy of each subhalo of the VL2 sample and the ratio between $v_{\rm max}(z=0)$ and $v_{\rm max}(z_{\rm infall})$, denoted by $v_{\rm peak}$.}
\label{energyvpeak}
\end{center}
\end{figure}

For this subhalo analysis, we use the Via Lactea II (VL2) simulation\footnote{The subhalo data set of VL2 simulation can be found here: www.itp.uzh.ch/~diemand/vl/data.html.} \cite{Diemandetal2008}, a high resolution $\Lambda$CDM realisation of a refined dark matter halo with a mass similar to the Milky Way. We define the binding energy of a subhalo as the sum of kinetic plus potential energy, divided by the mass of the subhalo \cite[][]{BinneyTremaine}, 
\begin{eqnarray}
e_{\rm orbit} &\equiv& K + U \nonumber \\
    &=& \frac{1}{2}\mathbf{v}^2 - \frac{G M}{r}\log(1 + \frac{r}{r_s}),
\end{eqnarray}
where $\mathbf{v}$ is the velocity of the subhalo with respect to the halo center and the main halo is fitted with an NFW mass distribution \cite{Navarroetal1997} with scale radius $r_s$ of 21.5 kpc. A subhalo is considered to be bound to the main halo if $e_{\rm orbit} < 0$. Furthermore, only subhalos are listed in the sample that (a) lie within the virial radius of the main halo and (b) have a maximum circular velocity at infall of $v_{\rm max}(z_{\rm infall}) \gtrsim 10$km/s. This reduces the original set to a sample of 953 objects.

Fig. \ref{energyvpeak} shows the median values of the ratio $v_{\rm max}(z=0) / v_{\rm peak}$ for these 953 subhalos binned by their orbital energy (red solid dots). The ratio $v_{\rm max}(z=0) / v_{\rm max}(z_{\rm infall})$ is a proxy for the subhalo mass loss since the time of accretion due to tidal stripping. The green shaded area shows the region between the $16^{\rm th}$ and $84^{\rm th}$ percentiles. A tight correlation is clearly visible: the deeper a subhalo orbits within the potential well of the main halo, the more tidal stripping it has undergone. This leads to a reduced value of the maximum circular velocity at present time relative to the peak value at infall. The black solid line in Fig. \ref{energyvpeak} is a simple regression with respect to the median, given by the following one parameter fitting form:
\begin{equation}\label{energyfit}
\frac{v_{\rm max}}{v_{\rm peak}}\big(e_{\rm orbit}\big) = 1.0 + \xi \cdot e_{\rm orbit}.
\end{equation}
The value of the fitting parameter is obtained using the Levenberg \& Marquart method, yielding $\xi = 2.89\cdot10^{-6}$ [km/s]$^{-2}$. The boundary condition of the above fitting form was chosen such that unbound subhalos ($e_{\rm orbit} = 0$) have no mass loss ($v_{\rm max} = v_{\rm peak}$). Equation \eqref{energyfit} now provides a simple way to convert the present day value of the maximum circular velocity to a quantity $v_{\rm peak}$, corresponding to the maximum circular velocity at time of accretion. However, we want to emphasize that this conversion has only been tested so far for a Milky Way sized cold dark matter halo.
\begin{figure}
\includegraphics[scale=0.38]{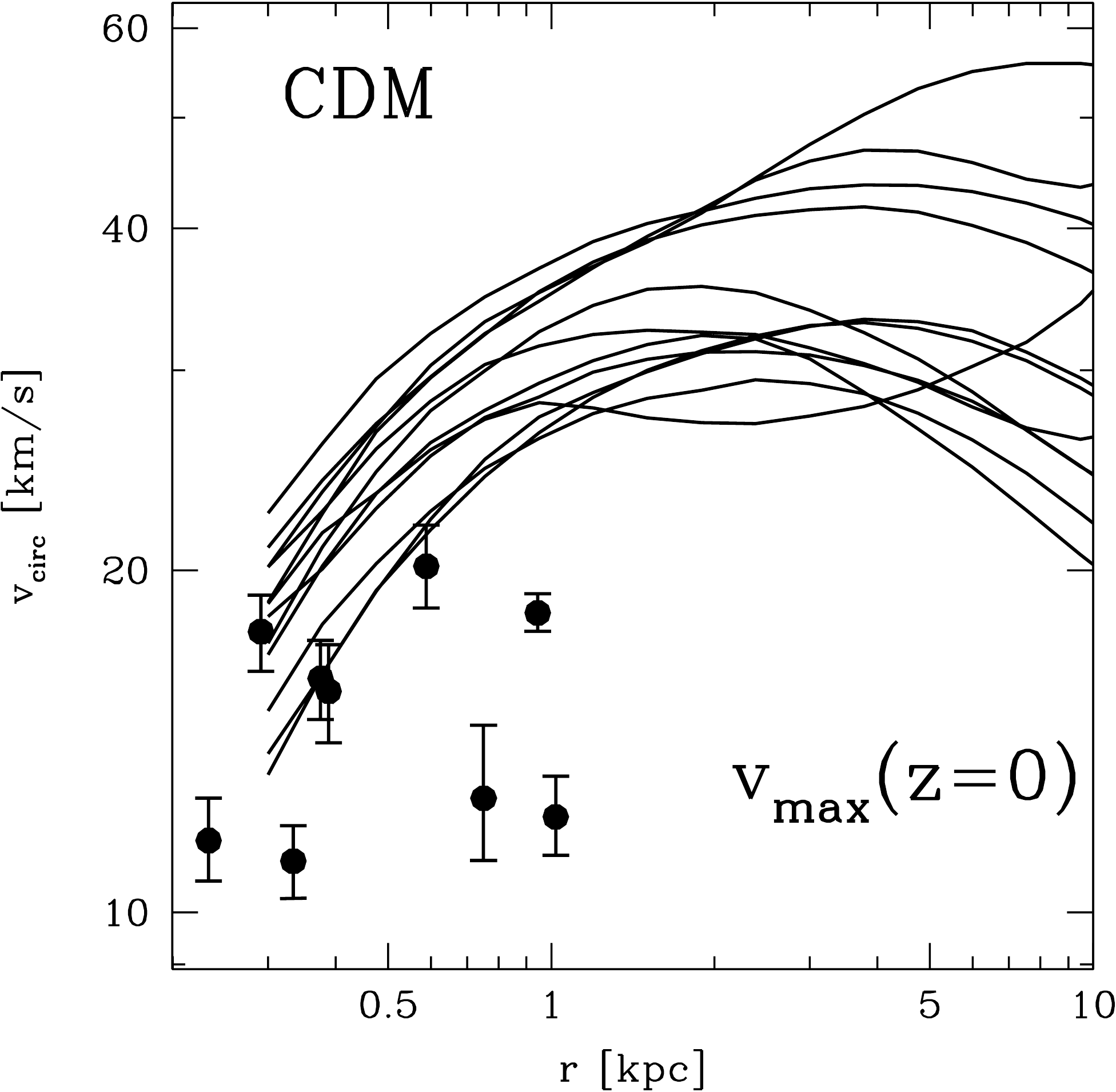}
\includegraphics[scale=0.38]{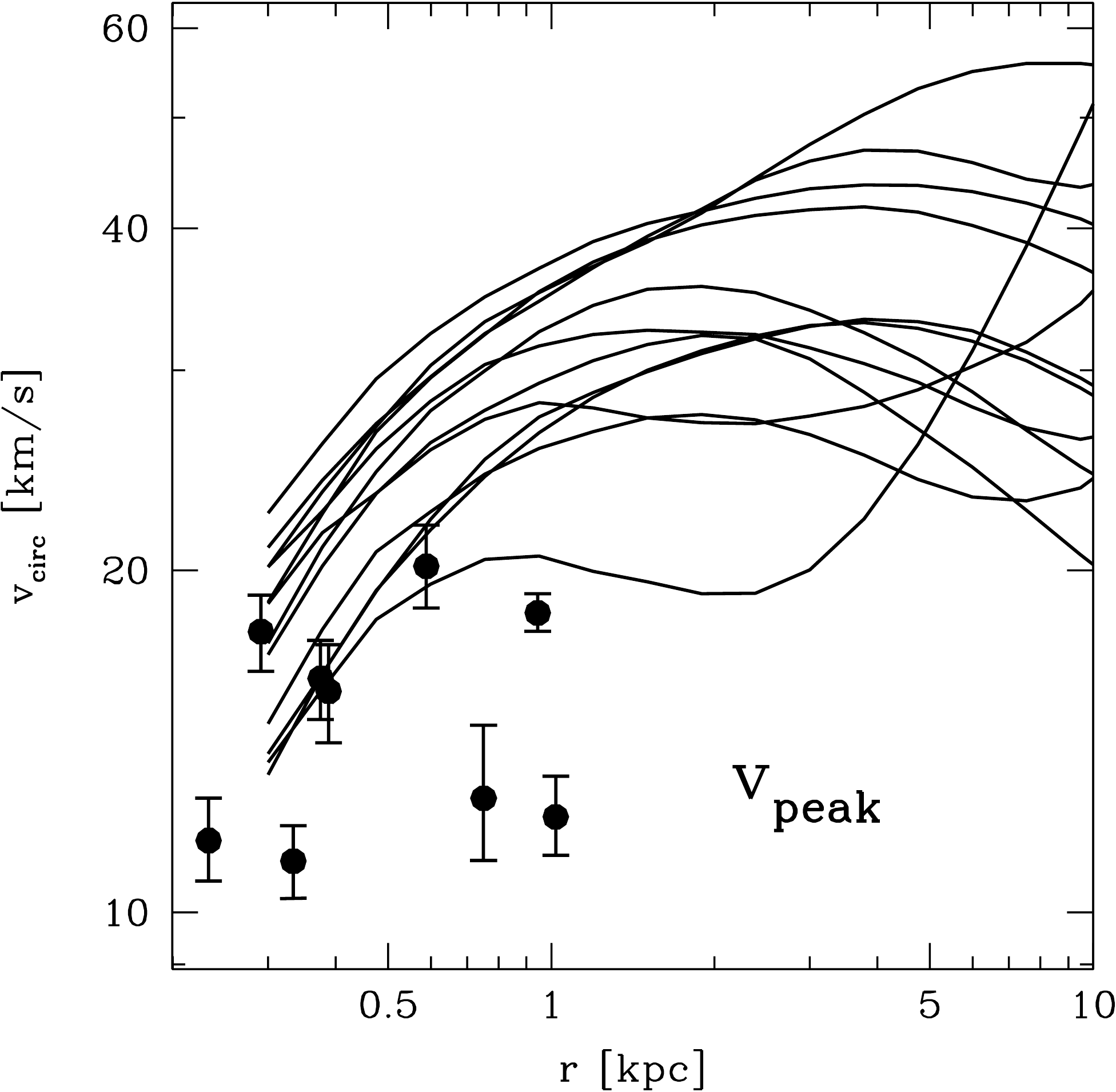}
\newline
\includegraphics[scale=0.38]{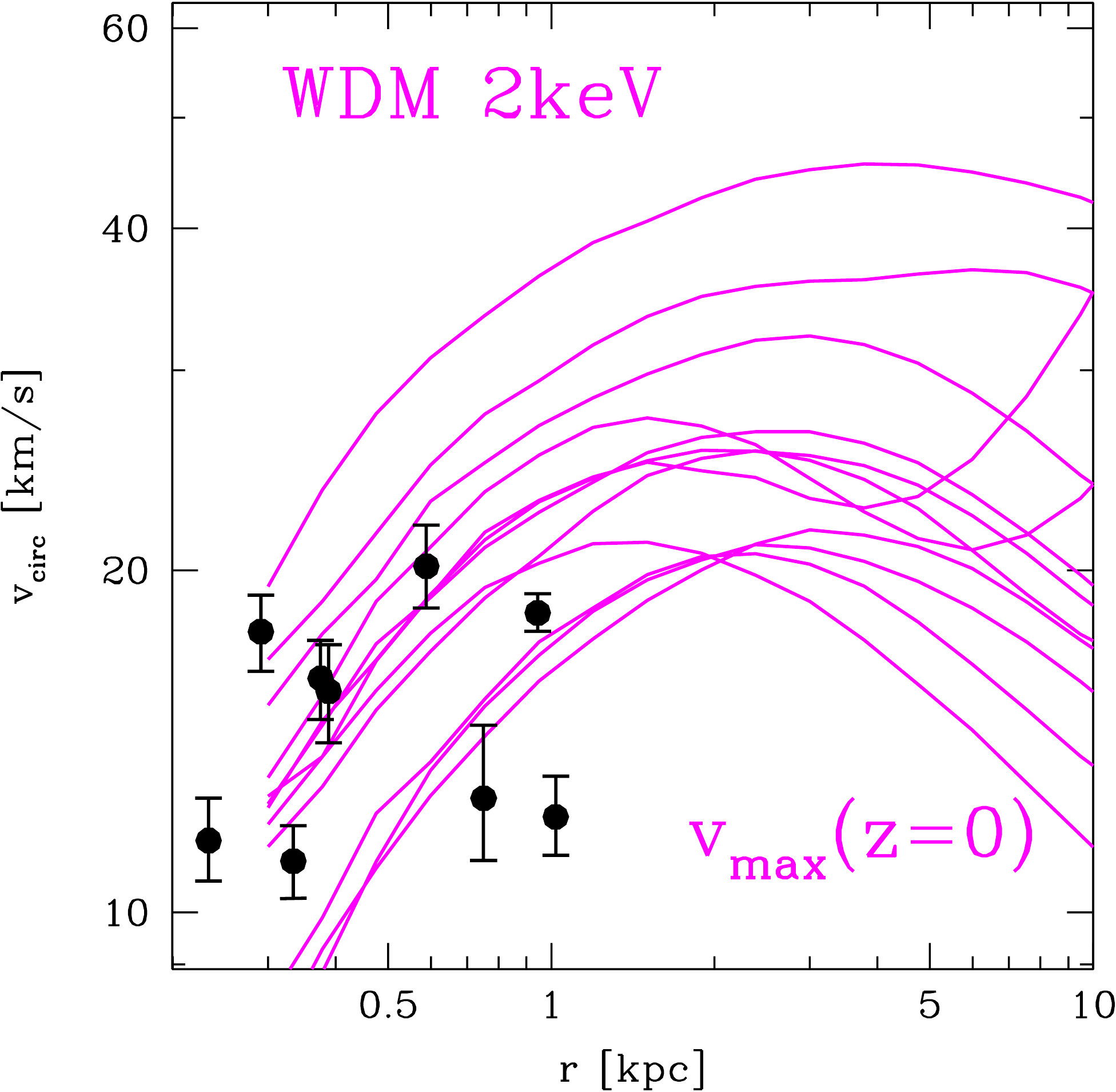}
\includegraphics[scale=0.38]{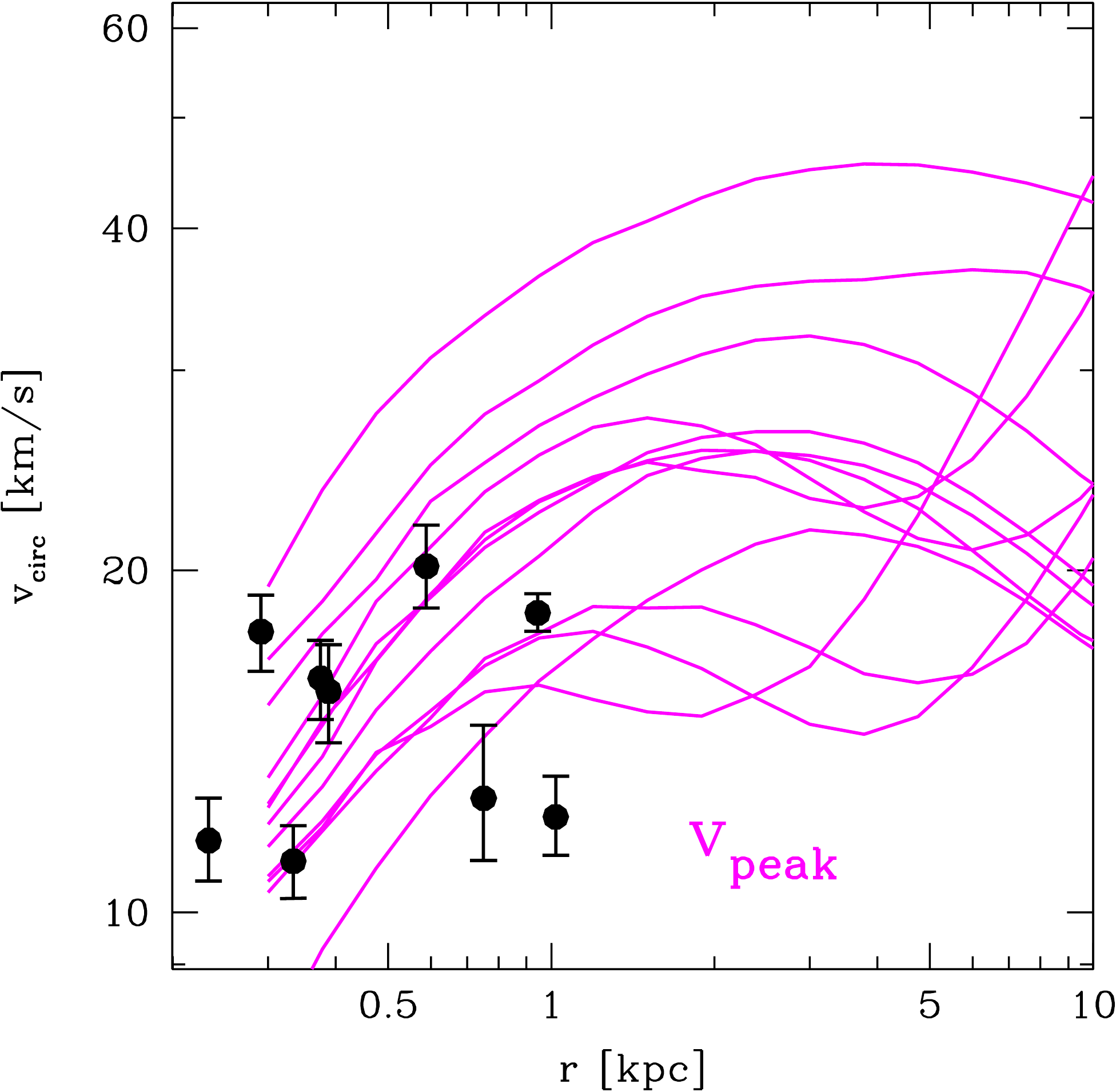}
\caption{Left panels: circular velocity curves fo the 12 most massive subhalos at redshift $z=0$ ($v_{\rm max}(z=0)$). Right panels: circular velocity curves of the 12 most massive subhalos at infall ($v_{\rm peak}$), calculated with our correction scheme (Eq. \eqref{energyfit}). Each is illustrated for the pure CDM (top) as well as the pure WDM (bottom) simulation. The data points refer to the 9 brightest satellite galaxies in the Milky Way.}
\label{vcirccorrection}
\end{figure}

The impact of this correction scheme is illustrated in Fig. \ref{vcirccorrection} for two of the simulations, CDM and WDM respectively. The left panels show the circular velocity curves of the 12 most massive subhalos at present time ($v_{\rm max}(z=0)$), whereas the right panels show the sample when the selection is made by $v_{\rm peak}$. Even if the selection methods do overlap to some extent, the right panels of Fig. \ref{vcirccorrection} unsheathe quite a few subhalos that have lost a lot of their mass since the time of accretion. A more detailed comparison between different subhalo selection methods is shown in \cite{Boylan-Kolchinetal2012}.


\begin{thebibliography}{99}

\bibitem{SDSScollaboration}
J.~K. Adelman-McCarthy et al., {\it The Fifth Data Release of the Sloan Digital Sky Survey}, ApJS, 172 (2007), 634

\bibitem{6dFcollaboration2009}
D.~H. Jones et al., {\it The 6dF Galaxy Survey: Final Redshift Release (DR3) and Southern Large-Scale Structures}, \mnras, 399 (2009), 683

\bibitem{Swatersetal2003}
R.~A. Swaters, B.~F. Madore, F.~C. van den Bosch \& M. Balcells, {\it The Central Mass Distribution in Dwarf and Low Surface Brightness Galaxies}, \apj, 583 (2003), 732

\bibitem{Gentileetal2009}
G. Gentile, B. Famaey, H. Zhao \& P. Salucci, {\it Universality of galactic surface densities within one dark halo scale-length}, Nature, 461 (2009), 627

\bibitem{Strigarietal2010}
L.~E. Strigari, C.~S. Frenk \& S.~D.~M. White, {\it Kinematics of Milky Way Satellites in a Lambda cold dark matter universe}, \mnras, 408 (2010), 2364

\bibitem{Walker2011}
M.~G. Walker \& J. Penarrubia, {\it A Method for Measuring (Slopes of) Mass Profiles of Dwarf Spheroidal Galaxies}, \apj, 742 (2011), 20

\bibitem{Agnello2012}
A. Agnello \& N.~W. Evans, {\it A Virial Core in the Sculptor Dwarf Spheroidal Galaxy}, \apj, 754 (2012), L39

\bibitem{Moore94}
B. Moore, {\it The Nature Of Dark Matter}, Nature, 370 (1994), 629

\bibitem{Diemandetal2005}
J. Diemand, M. Zemp, B. Moore, J. Stadel \& M.~C. Carollo, {\it Cusps in cold dark matter haloes}, \mnras, 364 (2005), 665

\bibitem{maccio2012}
A.~V. Macci\`o, S. Paduroiu, D. Anderhalden, A. Schneider \& B. Moore, {\it Cores in warm dark matter haloes: a Catch 22 problem}, \mnras, 424 (2012), 1105

\bibitem{Mooreetal1999c}
B. Moore, T. Quinn, F. Governato, J. Stadel, \& G. Lake, {\it Cold collapse and the core catastrophe}, \mnras, 310 (1999b), 1147

\bibitem{Klypinetal1999}
A. Klypin, A.~V. Kravtsov, O. Valenzuela, \& F. Prada, {\it Where Are the Missing Galactic Satellites?}, \apj, 522 (1999), 82

\bibitem{Boylan-Kolchinetal2011}
M. Boylan-Kolchin, J.~S. Bullock, \& M. Kaplinghat, {\it Too big to fail? The puzzling darkness of massive Milky Way subhaloes}, \mnras, 415 (2011), L40

\bibitem{Lovelletal2012}
M. Lovell, V. Eke, C. Frenk, L. Gao, A. Jenkins, T. Theuns, J. Wang, A. Boyarsky, \& O. Ruchayskiy, {\it The haloes of bright satellite galaxies in a warm dark matter universe}, \mnras, 420 (2012), 2318

\bibitem{Rashkovetal2012}
V. Rashkov, P. Madau, M. Kuhlen, \& J. Diemand, {\it On the Assembly of the Milky Way Dwarf Satellites and Their Common Mass Scale}, \apj, 745 (2012), 142

\bibitem{Tollerudetal2012}
E.~J. Tollerud, R.~L. Beaton, M. Geha, J.~S. Bullock, J.~S. Kalirai, E.~N. Kirby \& M. Boylan-Kolchin, {\it Kinematics Of M31 dSphs And Implications For LCDM}, AAS, 219 (2012), 201.04

\bibitem{Governatoetal2012}
F. Governato, A. Zolotov, A. Pontzen, C. Christensen, S.~H. Oh, A.~M. Brooks, T. Quinn, S. Shen \& J. Wadsley, {\it Cuspy no more: how outflows affect the central dark matter and baryon distribution in Lambda cold dark matter galaxies}, \mnras, 422 (2012), 1231

\bibitem{Bodeetal2001}
P. Bode, J.~P. Ostriker \& N. Turok, {\it Halo Formation in Warm Dark Matter Models}, \apj, 556 (2001), 93

\bibitem{Schneideretal2012}
A. Schneider, R.~E. Smith, A.~V. Macci\`o \& B. Moore, {\it Non-linear evolution of cosmological structures in warm dark matter models}, \mnras, 424 (2012), 684

\bibitem{Ellisetal1984}
J. Ellis, J.~S. Hagelin, D.~V. Nanopoulos, K. Olive \& M. Srednicki, {\it}, Nucl. Phys. B, 238 (1984), 453

\bibitem{Moroietal1993}
T. Moroi, H. Murayama, M. Yamaguchi, {\it}, Phys. Lett. B, 303 (1993), 289

\bibitem{Gorbunovetal2008}
D. Gorbunov, A. Khmelnitsky \& V. Rubakov, {\it }, J. High Energy Phys., 12 (2008), 55

\bibitem{DodelsonWidrow1994}
S. Dodelson \& L.~M. Widrow, {\it}, Physical Review Letters, 72 (1994), 17

\bibitem{ShaposhnikovTkachev2006}
M. Shaposhnikov \& I. Tkachev, {\it}, Phys. Lett. B, 639 (2006), 414

\bibitem{Shaposhnikov2007}
M. Shaposhnikov, {\it Dark Matter: The Case of Sterile Neutrino}, Conference Proceedings (2007), [astro-ph/0703673]

\bibitem{boyarsky2009}
A. Boyarsky, O. Ruchayskiy \& M. Shaposhnikov, {\it The Role of Sterile Neutrinos in Cosmology and Astrophysics}, Annual Review of Nuclear and Particle Science, 59 (2009), 191

\bibitem{Boyarskyetal2009a}
A. Boyarsky, J. Lesgourgues, O. Ruchayskiy \& M. Viel, {\it Lyman-alpha constraints on warm and on warm-plus-cold dark matter models}, JCAP, 05 (2009), 012

\bibitem{Boyarskyetal2009b}
A. Boyarsky, J. Lesgourgues, O. Ruchayskiy \& M. Viel, {\it Realistic sterile neutrino dark matter with keV mass does not contradict cosmological bounds}, Phys.Rev. Lett. 102 (2009), 201304

\bibitem{MaccioMixed}
 A.~V. Macci\`o, O. Ruchayskiy, A. Boyarsky \& J.~C. Munoz-Cuartas, {\it The inner structure of haloes in Cold+Warm dark matter models}, \mnras, 482 (2013), 882
 
\bibitem{Meiksin2007}
A.~A. Meiksin, {\it Physics of the Intergalactic Medium}, 2007, [astro-ph/0711.3358]

\bibitem{Anderhaldenetal2012}
D. Anderhalden, J. Diemand, G. Bertone, A.~V. Macci\`o \& A. Schneider, {\it The galactic halo in mixed dark matter cosmologies}, JCAP, 10 (2012), 047

\bibitem{Vieletal2005}
M. Viel, J. Lesgourgues, M.~G. Haehnelt, S. Matarrese \& A. Riotto, {\it Constraining warm dark matter candidates including sterile neutrinos and light gravitinos with WMAP and the Lyman-alpha forest}, \prd, 71 (2005), 063534

\bibitem{Stadel2001}
J. Stadel, {\it Cosmological N-Body Simulations and their Analysis}, PhD Thesis, University of Washington, U.S.A. (2001)

\bibitem{Bertschinger2001}
E. Bertschinger, {\it Multiscale Gaussian Random Fields and Their Application to Cosmological Simulations}, ApJS, 137 (2001), 1

\bibitem{Komatsuetal2011}
WMAP collaboration, E. Komatsu et al., {\it Seven-Year Wilkinson Microwave Anisotropy Probe (WMAP) Observations: Cosmological Interpretation}, ApJS, 192 (2011), 18

\bibitem{villa11}
F. Villaescusa-Navarro \& N. Dalal, {\it Cores and Cusps in Warm Dark Matter Halos}, JCAP, 03 (2011), 024

\bibitem{Diemandetal2006}
J. Diemand, M. Kuhlen \& P. Madau, {\it Early supersymmetric cold dark matter substructure}, \apj, 649 (2006), 1

\bibitem{Zempetal2012}
M. Zemp, O.~Y. Gnedin, N.~Y. Gnedin \& A.~V. Kravtsov, {\it The Impact of Baryon Physics on the Structure of High-redshift Galaxies}, \apj, 748 (2012), 54

\bibitem{Colin2008}
P. Colin, O. Valenzuela \& V. Avila-Reese, {\it On the Structure of Dark Matter Halos at the Damping Scale of the Power Spectrum with and without Relict Velocities}, \apj, 673 (2008), 203

\bibitem{DalalAnderhalden2013}
N. Dalal \& D. Anderhalden, 2013, in prep.

\bibitem{WangWhite}
J. Wang \& S.~D.~M. White, {\it Discreteness effects in simulations of hot/warm dark matter}, \mnras, 380 (2007), 93

\bibitem{DiemandMoore2011}
J. Diemand \& B. Moore, {\it The Structure and Evolution of Cold Dark Matter Halos}, Advanced Science Letters, 4 (2011), 297

\bibitem{Kuhlenetal2012}
M. Kuhlen, M. Vogelsberger \& R. Angulo, {\it Numerical Simulations of the Dark Universe: State of the Art and the Next Decade}, eprint arXiv:1209.5745 (2012)

\bibitem{Stoehretal2002}
F. Stoehr, S.~D.~M. White, G. Tormen \& V. Springel, {\it The satellite population of the Milky Way in a LCDM universe}, \mnras, 335 (2002), L84

\bibitem{Penarubbiaetal2008}
J. Pe\~{n}arrubia, A.~W. McConnachie \& J.~F. Navarro, {\it The Cold Dark Matter Halos of Local Group Dwarf Spheroidals}, \apj, 672 (2008), 904

\bibitem{Bullocketal2000}
J.~S. Bullock, A.~V. Kravtsov \& D.~H. Weinberg, {\it Reionization and the Abundance of Galactic Satellites}, \apj, 539 (2000), 517

\bibitem{Kravtsovetal2004}
A.~V. Kravtsov, O.~Y. Gnedin \& A.~A. Klypin, {\it The Tumultuous Lives of Galactic Dwarfs and the Missing Satellites Problem}, \apj, 609 (2004), 482

\bibitem{Bushaetal2010}
M.~T. Busha, M.~A. Alvarez, R.~H. Wechsler, T. Abel \& L.~E. Strigari, {\it The Impact of Inhomogeneous Reionization on the Satellite Galaxy Population of the Milky Way}, \apj, 710 (2010), 408

\bibitem{Belokurovetal2007}
V. Belokurov, et al., {\it Cats and Dogs, Hair and a Hero: A Quintet of New Milky Way Companions}, \apj, 654 (2007), 111b

\bibitem{Koposovetal2008}
S. Koposov et~al. 2008, {\it The Luminosity Function of the Milky Way Satellites}, \apj, 686, 279 (SK08)

\bibitem{Strigarietal2007}
L. Strigari, J.~S. Bullock, M. Kaplinghat, J. Diemand, M. Kuhlen \& P. Madau, {\it Redefining the Missing Satellites Problem}, \apj, 669 (2007), 676

\bibitem{Polisensky2011}
E. Polisensky \& M. Ricotti, {\it Constraints on the dark matter particle mass from the number of Milky Way satellites
}, Physical Review D, 83 (2011), 4

\bibitem{Wolfetal2010}
J. Wolf, G.~D. Martinez, J.~S. Bullock, M. Kaplinghat, M. Geha, R.~R. Mu$\tilde{\rm n}$oz, J.~D. Simon \& F.~F. Avedo, {\it Accurate masses for dispersion-supported galaxies}, \mnras, 183 (2010), 1220

\bibitem{Strigarietal2007}
L. Strigari, J.~S. Bullock, M. Kaplinghat, J. Diemand, M. Kuhlen \& P. Madau, {\it Redefining the Missing Satellites Problem}, \apj, 669 (2007), 676

\bibitem{Madauetal2008}
P. Madau, J. Diemand \& M. Kuhlen, {\it Dark Matter Subhalos and the Dwarf Satellites of the Milky Way}, \apj, 679 (2008), 1260

\bibitem{Maccioetal2010}
A.~V. Macci\`o, X. Kang, F. Fontanot, R.~S. Sommerville, S. Koposov \& P. Monaco, {\it Luminosity function and radial distribution of Milky Way satellites in a LCDM Universe}, 2010, \mnras, 402, 1995

\bibitem{Munozetal2009}
J.~A. Mu$\tilde{\rm n}$oz, P. Madau, A. Loeb \& J. Diemand, {\it Probing the epoch of reionization with Milky Way satellites}, \mnras, 400 (2009), 1593

\bibitem{Diemandetal2008}
J. Diemand, M. Kuhlen, P. Madau, M. Zemp, B. Moore, D. Potter \& J. Stadel, {\it Clumps and streams in the local dark matter distribution}, Nature, 454 (2008), 735

\bibitem{Klypinetal2011}
A.~A. Klypin, S. Trujillo-Gomez \& J. Primack, {\it Dark Matter Halos in the Standard Cosmological Model: Results from the Bolshoi Simulation}, \apj, 740 (2011), 102

\bibitem{Tollerud2010}
E.~J. Tollerud, {\it Hundreds of Milky Way Satellites and a Fundamental Curve Connecting Dark Matter Halos to Galaxies}, AIP Conference Proceedings, 1240 (2010), 425

\bibitem{Boylan-Kolchinetal2012}
M Boylan-Kolchin, J.~S. Bullock \& M. Kaplinghat, {\it The Milky Way's bright satellites as an apparent failure of LCDM}, \mnras, 422 (2012), 1203

\bibitem{PurcellZentner2012}
C.~W. Purcell \& A.~R. Zentner, {\it Bailing Out the Milky Way: Variation in the Properties of Massive Dwarfs Among Galaxy-Sized Systems}, eprint arXiv:1208.4602 (2012)

\bibitem{Gnedinetal2010}
O.~Y. Gnedin, W.~R. Brown, M.~J. Geller \& S.~J. Kenyon, {\it The Mass Profile of the Galaxy to 80 kpc}, ApJL, 720 (2010), L108

\bibitem{LiWhite}
Y.~S. Li \& S.~D.~M. White, {\it Masses for the Local Group and the Milky Way}, \mnras, 384 (2008), 1459

\bibitem{Watkinsetal2010}
L.~L. Watkins, N.~W. Evans \& J.~H. An, {\it The masses of the Milky Way and Andromeda galaxies}, \mnras, 406 (2010), 264

\bibitem{Bushaetal2011}
M.~T. Busha, P.~J. Marshall, R.~H. Wechsler, A. Klypin \& J. Primack, {\it The Mass Distribution and Assembly of the Milky Way from the Properties of the Magellanic Clouds}, \apj, 743 (2011), 40

\bibitem{Xueetal2008}
X.~X. Xue et~al., {\it The Milky Way's Circular Velocity Curve to 60 kpc and an Estimate of the Dark Matter Halo Mass from Kinematics of $\sim$2400 SDSS Blue Horizontal-Branch Stars}, \apj, 684 (2008), 1143

\bibitem{Kafleetal2012}
P.~R. Kafle, S. Sharma, G.~F. Lewis, J. Bland-Hawthorn, {\it Kinematics of the Stellar Halo and the Mass Distribution of the Milky Way Using Blue Horizontal Branch Stars}, \apj, 761 (2012), 98

\bibitem{Boylan-Kolchinetal2010}
M. Boylan-Kolchin, V. Springel, S.~D.~M. White \& A. Jenkins, {\it There's no place like home? Statistics of Milky Way-mass dark matter halos}, \mnras, 406 (2010), 896

\bibitem{Vogelsbergeretal2012}
M. Vogelsberger, J. Zavala \& A. Loeb, {\it Subhaloes in self-interacting galactic dark matter haloes}, \mnras, 423 (2012), 3740

\bibitem{Diemandetal2007}
J. Diemand, M. Kuhlen \& P. Madau, {\it Evolution of Dark Matter Subhalos}, AAS Meeting 210, 39 (2007), 144

\bibitem{Parryetal2012}
O.~H. Parry, V.~R. Eke, C.~S. Frenk \& T. Okamoto, {\it The baryons in the Milky Way satellites}, \mnras, 419 (2012), 3304

\bibitem{WadepuhlSpringel}
M. Wadepuhl \& V. Springel, {\it Satellite galaxies in hydrodynamical simulations of Milky Way sized galaxies}, \mnras, 410 (2011), 1975

\bibitem{Rochaetal2012}
M. Rocha, A.~H.~G. Peter \& J.~S. Bullock, {\it Infall times for Milky Way satellites from their present-day kinematics}, \mnras, 425 (2012), 231

\bibitem{BoylanetalLeo1}
M. Boylan-Kolchin, J.~S. Bullock, S.~T. Sohn, G. Besla \& R.~P. van der Marel, {\it The Space Motion of Leo I: The Mass of the Milky Way's Dark Matter Halo}, [astro-ph/1210.6046]

\bibitem{BinneyTremaine}
J. Binney \& S. Tremaine, {\it Galactic Dynamics}, Princeton University Press (2008)

\bibitem{Navarroetal1997}
J.~F. Navarro, C.~S. Frenk \& S.~D.~M. White, {\it A Universal Density Profile from Hierarchical Clustering}, \apj, 490 (1997), 493

\end{thebibliography}
\end{document}